\clearpage{}%

\documentclass[times]{cpeauth}

\usepackage{moreverb}

\newcommand\BibTeX{{\rmfamily B\kern-.05em \textsc{i\kern-.025em b}\kern-.08emT\kern-.1667em\lower.7ex\hbox{E}\kern-.125emX}}

\usepackage{graphicx}
\usepackage{url} 
\usepackage{color} 
\usepackage{enumerate}
\usepackage{longtable} 
\usepackage{textcomp} %
\usepackage{setspace} \usepackage{lscape} %
\usepackage{paralist}

\definecolor{orange}{rgb}{1.0,0.3,0.0} \definecolor{violet}{rgb}{0.75,0,1}
\definecolor{darkgreen}{rgb}{0,0.6,0} \definecolor{cyan}{rgb}{0.2,0.7,0.7}
\definecolor{blueish}{rgb}{0.2,0.2,0.8}

\newif\ifdraft 
\drafttrue 
\ifdraft 
\newcommand{\katznote}[1]{{\textcolor{magenta} { ***Dan: #1 }}} 
\newcommand{\rananote}[1]{{\textcolor{cyan} { ***Omer: #1 }}} 
 
\newcommand{\note}[1]{ {\textcolor{red} { #1 }}}
\newcommand{\jhanote}[1]{ {\textcolor{blue} { ***Shantenu: #1 }}}
\newcommand{\jonnote}[1]{ {\textcolor{red} { ***Jon: #1 }}}
\newcommand{\alnote}[1]{ {\textcolor{darkgreen} { ***Andre: #1 }}}
\newcommand{\nchnote}[1]{ {\textcolor{orange} { ***Neil: #1 }}}
\newcommand{\ysnote}[1]{ {\textcolor{violet} { ***Yogesh: #1 }}}
 
\else 
\newcommand{\katznote}[1]{}
\newcommand{\rananote}[1]{} 
\newcommand{\note}[1]{} 
\newcommand{\jhanote}[1]{}
\newcommand{\jonnote}[1]{} 
\newcommand{\alnote}[1]{} 
\newcommand{\nchnote}[1]{}
\newcommand{\ysnote}[1]{} 
\fi

\usepackage[pdftex,colorlinks=true, linkcolor=blue,citecolor=blue]{hyperref} 
\clearpage{}%

\begin{document}
\runningheads{S.~Jha et al.}{Understanding Applications and Infrastructure for D3 Science}

\title{Introducing Distributed Dynamic Data-intensive (D3) Science:
  Understanding Applications and Infrastructure}

\author{Shantenu Jha\affil{1}\corrauth\, Daniel S. Katz\affil{2},
  Andre~Luckow\affil{1}, \\ Omer Rana\affil{3}, Yogesh Simmhan\affil{4}, Neil Chue
  Hong\affil{5}}

\address{\affilnum{1}Rutgers University\break
\affilnum{2}University of Chicago \& Argonne National Laboratory\break
\affilnum{3}Cardiff University\break
\affilnum{4}Indian Institute of Science\break
\affilnum{5}University of Edinburgh\break
}

\corraddr{Rutgers University, NJ, USA. E-mail: shantenu.jha@rutgers.edu}


\begin{abstract}
  A common feature across many science and engineering applications is the
  amount and diversity of data and computation that must be integrated to yield
  insights. Data sets are growing larger and becoming distributed; and their
  location, availability and properties are often time-dependent. Collectively, these
  characteristics give rise to dynamic distributed data-intensive
  applications. While ``static'' data applications have received significant
  attention, the characteristics, requirements, and software systems for the
  analysis of large volumes of dynamic, distributed data, and data-intensive applications have
  received relatively less attention. This paper surveys several representative dynamic
  distributed data-intensive application scenarios, provides a common conceptual
  framework to understand them, and examines the infrastructure used in support
  of applications.

\end{abstract}
\maketitle

\section{Introduction: Context, Scope and Outline}

The landscape of scientific and enterprise computing is being fundamentally
altered by prodigious volumes of data.  
This massive data generation can be naturally distributed because they are being created in a set of distributed
locations, rather than in one place. At the same time, data can also be
transported to be processed in multiple, distributed locations, rather than
being processed near their source, or be brought together for centrally
processing at a particular location. These two orthogonal dimensions lead to a
number of patterns of data generation and processing being present in real
applications, either for reasons of performance or for collaborative analytics.
Further, the variation of data production rates, data source, and destination
makes temporal variations in data properties increasingly important. With an
increase in the importance of distributed and temporal properties of data,
systems and infrastructure aspects such as resource scheduling, data placement,
and transfer decisions that were statically determinable at small scales emerge
as dynamic and important considerations at large scales of data and
distribution.

The focus of this paper is on a subspace of the large set of problems associated
with data-intensive sciences, namely that which is characterized by {\it both}
distributed and dynamic aspects. We introduce the concept of Dynamic Distributed
Data-intensive science and refer to this subspace of data-intensive science as
D3 science.

The analytical space spanned by the three ``D''s of dynamic, distributed,
and data-intensive provides a common vocabulary and effective framework to analyze
applications and infrastructure alike.  Subjecting a range of otherwise distinct
applications to a common analytical framework provides an opportunity to search
for similarities across applications and understand core differences.  The
diverse D3 scenarios point to the need for a careful understanding of D3
characteristics as well as sophisticated, extensible, and skillfully architected
infrastructure.  A common framework thus informs infrastructure scientists and
cyberinfrastructure experts where effort is likely to yield maximum returns.
The advantages of providing a common but extensible vocabulary that can be used
by the entire community---application end-users, developers and infrastructure
scientists alike---in an emerging field cannot be overstated. We believe that
such a vocabulary and terminology has been missing from the practice of
distributed cyberinfrastructure in general, with negative consequences.

\subsection{Scope}

One motivation of this work is to understand the dynamic properties
and the distribution of applications, and the complex interplay
between them that arises as a consequence of scale.  Additionally, we
are motivated by an attempt to understand infrastructure---existing
capabilities, trends, and limitations---available to support the complexity,
challenges and characteristics of dynamic and distributed data at
scale.

We would like the reader to keep in mind some points regarding context and scope:

\begin{itemize}

\item This {\it survey} is not meant to be complete, either in the scope of domains
  covered or the types of infrastructure and application surveyed.  It
  represent the interests and experience of the authors. The specific
  requirements and barriers of applications cited in this paper are
  representative of the time at which the applications were surveyed
  (2012-2013); this is also true of the quantitative analysis of applications.

\item The focus is generally on applications and infrastructure that arise from science
  and engineering projects in academia in general; this is a
  guiding principle.  While we are cognizant of enterprise applications and
  infrastructure, and recognize the impressive advances made therein, we also
  acknowledge that the factors such as scale, distribution, open-source, and
  interoperability influence the solutions employed and infrastructures used; it
  is important to therefore focus on problems that emerge in the space
  defined by these ``academic constraints''.

\item Almost by definition, any attempt to capture the state-of-the-art of such
  a fast-changing and emerging domain is bound to be fraught with limitations 
  and issues of scoping.  A complete and rigorous survey is bound to be 
  obsolete by
  the time it emerges; so given the rate of change, we asked what could be
  done. What we present is a partial analysis motivated by a set of existing and
  active D3 applications, their trials and tribulations, successes and
  solutions. We believe there is merit in capturing the state-of-play as is,
  deriving a research agenda and set of specific yet crosscutting issues, and
  presenting these to the community.

\item Distribution can occur on many levels.  In this paper, we use {\em locally
    distributed} to indicate that data generation and storage, or computation 
	over it, occurs on multiple nodes within one data center.  
  {\em Geographically distributed} describes the distribution
  of data generation, storage and/or compute across multiple data centers.  
  Also, our focus is
  primarily on infrastructure considerations at  large scale, as searching
  for common solutions and patterns at  large scale is likely to be more
  fruitful than at small scales where customized tools and optimization of
  low-level systems features is likely to return greater yield.
  
 \item We use the term dynamism to refer to the spatio-temporal variability of 
 data (e.\,g.\ data rates and formats), applications (e.\,g., workflows that 
 depend on data values) and infrastructures (e.\,g.\ availability).

\end{itemize}

\subsection{Outline of this paper}

The remainder of this paper consists of
the application scenarios (\S\ref{sec:scenarios}),
a discussion of the aspects of distribution and dynamism in the
applications (\S\ref{sec:distdyndata}),
a description of the software infrastructure required for D3 applications
(\S\ref{sec:infrastructure}),
analysis and characterization of the applications and
conclusions (\S\ref{sec:conclusions}). Each of these sections is briefly described below:

\begin{itemize}

\item Section~\ref{sec:scenarios} provides a description of thirteen D3 applications based on our survey. For each application, both the {\it dynamic} and {\it distributed} aspects are described. The methodology used to carry out the survey is described in the Appendix.

\item Based on the application survey, we derive a set of  definitions for {\it Distribution} and {\it Dynamism} in \S\ref{sec:distdyndata}. Note that these definitions are based on the application survey. A variety of definitions of these concepts already exist in computer science literature. Our intention is
to provide context for the use of these terms with reference to our survey. The distributed and dynamic characteristics of each application surveyed in \S\ref{sec:scenarios} is summarized in this section.

\item In \S\ref{sec:infrastructure}, we subsequently describe the software/systems infrastructure used to support the distributed and dynamic characteristics of applications in \S\ref{sec:scenarios}. We outline support for data management, analysis, and coordination, providing examples of actual cyberinfrastructure made use of by our surveyed applications for these purposes.

\item In \S\ref{sec:conclusions}, we bring together key observations based on the analysis of applications, their distributed and dynamic characteristics, and infrastructure use. An attempt is made to compare these applications and suggest the need for an ``architecture for D3 science".

\end{itemize}

\textit{Guidance for the reader}: Readers who are generally knowledge about distributed
applications will likely feel comfortable reading the paper from start to end.
Readers with less distributed applications knowledge may prefer to read
\S\ref{sec:distdyndata} and \S\ref{sec:infrastructure} before \S\ref{sec:scenarios},
or might want to read \S\ref{sec:scenarios} first, but jump to \S\ref{sec:distdyndata} and \S\ref{sec:infrastructure} when they find terms which which they are unfamiliar.
Unfortunately, we do not feel that there is a single ordering of the information in this
paper that is best for all readers.

\section{D3 Application Scenarios \label{sec:scenarios}}

Starting with the experience of the attendees of the Dynamic Distributed Data Programming Abstractions and Systems (3DPAS) workshops, we have
assembled a set of D3 ``applications'' that we describe in this section.  

The term ``application'' is often overloaded in literature and the computational
science discourse.  Sometimes the term is a reference to a standalone,
independent executable (e.g., a Molecular Dynamics simulation kernel such as
AMBER or Gromacs). Sometimes it refers to science problem to which
the executable is put to use (e.g., computing the free-energy of binding energy of
a drug candidate), but sometimes it is also used to reference the end-to-end
workflow (i.e., a multi-stage computational process requiring a sequence of
possibly distinct simulations and executables).  To add chaos to the confusion,
applications are sometimes also as a proxy for projects (e.g., the ATLAS
project), where the project could itself possibly be a series of
independent applications, with each application in turn a reference to a defined
science objective achieved using a set of executables.

In addition to the aforementioned hierarchy and granularity to which the term
application has been applied, the class of applications in any level differ
widely in scale, sophistication and type as well different objective and
objects.  Thus, we warn the reader of the inevitable befuddlement that the
casual and non-contextual use of the word application can cause.

Against this backdrop, we eschew a formal classification of application types,
however, to facilitate understanding we categorize applications into two
categories. The first is that of {\em traditional applications}, where
a program (or set of programs) is developed independently by a user or project to try to
answer a science question (or set of related questions). 
A single standalone computer program, a distributed application, or a
complex workflow could fall under the category of a {\em traditional
application}.  Applications in this paper that are in this category are NGS (next generation
sequencing) Analytics (\S\ref{bioSilvia}), CMB (cosmic microwave background)
(\S\ref{astroJulian}), Fusion (\S\ref{fusionScott}), Industrial Incident
Notification and Response (\S\ref{Kees}), MODIS (moderate resolution imaging
spectroradiometer) Data Processing (\S\ref{modisKeith}), and Distributed Network
Intrusion Detection (\S\ref{detectionJon}).

Another group of applications are {\em infrastructural applications}.
Applications of this type are collaborative, in that they depend on a context
that is often agreed upon by a community, most often in terms of how data is
stored by the community and how it is accessed, as well as, more often than not,
a weak consensus on the infrastructure to be used.  This allows different
sub-applications (which themselves are full-fledged ``applications'', but with dependencies on the collaborative infrastructure) to focus on
different stages, such as generating data, or processing data.  Answering a
science question now may involve a set of applications that need to be run in
series, perhaps in different stages, that may be run by different groups that do
not frequently interact.  In this paper, examples of this type of application include
ATLAS/WLCG (a
toroidal large hadron collider apparatus/worldwide large hadron collider
computing grid, \S\ref{WLCGSteve}), LSST (large synoptic survey telescope,
\S\ref{astroAdam}), SOA (service-oriented architecture) Astronomy
(\S\ref{astroSOAAdam}), Sensor Network Application (\S\ref{sensorSimon}),
Climate (\S\ref{climateDan}), Interactive Exploration of Environmental Data
(\S\ref{envJon}), and Power Grids (\S\ref{power}).

In other words, multiple infrastructural applications are used together in
particular science domains.  Furthermore, infrastructural applications involve
more than just the end user (scientist), and typically require exerting control
at lower layers, i.e., ``programming the system (infrastructure)'', as opposed
to a traditional applications where end user effort and control is often
confined to a well-defined application kernel and their communication and
coordination.  Also, for some {\em traditional applications} (e.g., NGS) the
number of users of an instance of the application is typically small, i.e., the
predominant use is by the long-tail of science. In contrast, some
infrastructural applications are correlated with ``big science'' projects.

Some of the applications we describe are really groups of
applications, meaning that we are grouping together a number of
independent programs, written by different authors, that are trying to
answer the same type of science question, where a single user
will only run one of them at a time.  %
These codes and their developers may either be
competing or collaborating, and sometimes both, informally known as
collabetition. Specifically, this is the case for NGS Analytics
(\S\ref{bioSilvia}), SOA Astronomy (\S\ref{astroSOAAdam}), and
Interactive Exploration of Environmental Data (\S\ref{envJon}).  (Note
that this issue is orthogonal to traditional vs. infrastructural applications:
NGS Analytics is traditional, where SOA Astronomy and Interactive
Exploration of Environmental Data are infrastructural.)  In
each of these examples, we have chosen one specific application that
is meant to represent a possibly large number of comparable
applications.

\begin{figure}[htbp]
    \centering
        \includegraphics[width=1.0\textwidth]{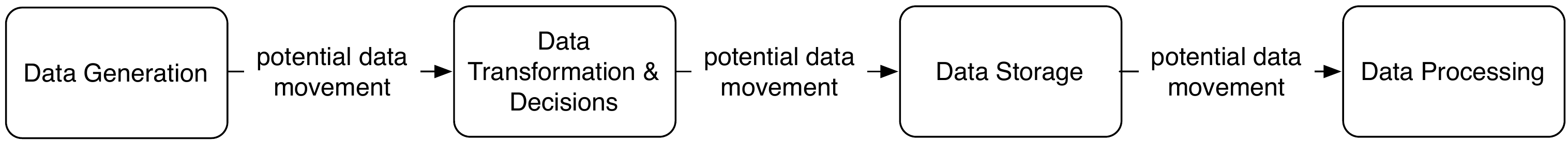}
        \caption{Application Stages: The work in data-intensive
        applications can be put into four common application stages:
		data generation, transformation \& decisions, storage, and processing, though not all
		applications have all stages.
                \label{fig:figures_application-stages}}
\end{figure}

The discussions in the workshops and our analysis of the applications
led us to examine the use of data in different parts of the
applications, 
and informed the terminology we use in the rest of the paper.
We have split the presentation of information on each scenario into three aspects: 
{\em big data aspects}, concerned with the number and volume of data; 
{\em distributed aspects}, concerned with the way data generation, storage and processing are separated; and 
{\em dynamic aspects}, concerned with the spatio-temporal variability of the data or application.
Doing this has enabled us to identify some common patterns
across applications.  For example, applications that involve
streaming data collect data from sensors, transform and
filter the data, store the results, and then later process (analyze)
the stored results.  In general, we think that all of the applications
we have studied can be mapped to a set of stages or phases.  This is
illustrated in \figurename~\ref{fig:figures_application-stages}.

The question of where the application, as opposed to the entire
system, %
actually starts
is somewhat difficult.  In this paper, we have decided not to treat
hardware data sources (e.g., sensors, telescopes) as part of the
applications.  Therefore, these applications generally start with the
data movement that follows the actual data generation.

Unlike the sensor applications, another set of applications generate
data computationally, for example, from a model or simulation.  These
applications then do start with data generation, and they often have
just three stages, where data is generated in the first stage, stored
in a second stage, and then processed in a third stage (i.e., the
transformation \& decisions stage in
\figurename~\ref{fig:figures_application-stages} does not occur).

Other
applications may operate on stored data (i.e., they only contain the
processing stage and its associate data movement).

\subsection{Next Generation Sequencing (NGS) Analytics \label{bioSilvia}}

The (grand) challenge in bioinformatics is the need to provide support
for the analysis of genome sequence data that is becoming available
due to the abundance and high-throughput capability of the
next-generation sequencing devices. The rate of growth of data from
the NGS machines humbles Moore's law; it has increased by more than
three orders of magnitude in barely a decade.

There are many different types and components of analyses of NGS data.
However, fundamentally important to all types is the ability to
efficiently and effectively map/align the short reads to a reference
genome.

Alignment underpins multiple scientific questions such as comparing
metagenomic (DNA sequence) data sets against each other, which could
be used to distinguish human disease biomarkers among different
individuals, determine different soil (or water or air or ...)
microbial composition from different samples, etc.

This subsection describes a set of {\em traditional applications} that
are used in the processing of next generation sequencing data.  The
applications have two of the possible stages: data storage and data
processing.

\subsubsection*{Application description}

There are multiple tools (programs) that can perform the mapping/alignment
phase, e.g., BWA, BFAST, Bowtie, MAQ, etc. Each has its own strength and
its own native data support/capability; they also implement
different alignment algorithms. Thus, based upon different design
objectives and trade-offs, they typically have different constraints
and efficiencies.

Several of these tools are also composed together as a linear pipeline
for performing sequencing and analysis. The pipeline may go through
alignment, SNP calling, haplotyping, and so on, with different tools
used in each stage. These pipelines can often be performed
independently on each chromosome to provide simple parallelism.

Some NGS analytics are not good fits for core-rich and memory-poor modern
architectures. Some problems (e.g., comparing k-mers across metagenomic data
sets) need terabyte-scale memory, not thousands of cores.~\cite{ngs-gap}

Individual data sets, which will be terabyte-scale, are generated by
NGS machines at a variety of locations. There could be many thousands
of such sets to compare. Having them all local may not be feasible
(the transfer time from where they are generated may be prohibitive).

\subsubsection*{Big data aspects}

Individual data sets will be terabyte-scale. For example, for a
standard genomic sequence, there are billions of short (35-100 bases)
reads of DNA.  And there could be many thousands of such sets to
compare for a given problem.  Getting more accuracy for sequencing may
require multiple coverage (e.g. 10x), that can cause sequence sizes to
increase linearly.  This makes NGS data management fundamentally a big
data problem.  NGS analytics, in addition to being a big data problem,
is also a computationally demanding and distributed computing problem.
The computational demands arise from the often complex and intensive
analysis that has to be performed on data, which in turn arise from
algorithms that are designed to account for repetitions, errors and
incomplete information.

\subsubsection*{Distributed aspects}

The type of NGS research being conducted can distinguish the
distributed properties of the application. Researcher groups that
own sequencing machines need to deal with large dataset sizes that
need to be re-sequenced and reduced for further analysis. Other
researchers may analyze large catalogs of sequenced genome databases
are maintained, and perform pattern matching. Further, clinical may be
interested in in-depth analysis of an individual sequence for disease
diagnosis or visualization.

The distributed computing aspects arise at multiple levels: for
example, the simple act of having to move data from source
(generation) to the destination where computing (analysis) will occur
is a challenge due to the large volumes involved. Trade-offs exists
between the cost/challenges in distributing data versus I/O
saturation or memory bottlenecks.

When coupled with the compute intensive nature of the problem, it soon
emerges that a fundamental challenge is not only whether to
distribute, but where to distribute, what to distribute (should the
computing move to the data, or the data move to the compute), and how
to distribute (what tools and infrastructure to use).

\subsubsection*{Dynamic aspects}

The dynamic challenges of NGS data are more subtle; the data
themselves are not dynamic -- the sequences are acquired once and
analyzed many times. However the execution of these applications on
distributed infrastructure have dynamic aspects when optimized
resource usage is considered. For example, workload decomposition and
distribution must be determined dynamically in order to optimally use
resources, e.g., compute resource selection can be based on data
location, processing profile and/or network capacity.

\subsubsection*{Other important issues}

There are several toolkits that have emerged for NGS providing
incrementally better algorithms for sequencing and analysis
\cite{SAMtools, BioJava, BioPerl, .NETBio}.  However, as a consequence
of the broad range of infrastructural requirements, there isn't
currently a common ``standard'' cyberinfrastructure used for NGS
analytics; researchers use what is easily available to them, or try to
use COTS infrastructure. Historically, leaders in the field have
developed infrastructure for the larger community to build-upon, e.g.,
QCDOC or MD-GRAPE. But given the broad range of requirements, a
hardware-only solution is unlikely for a broad range of problem
instances, and a combination of hardware and software approaches will
be employed. This reinforces the need for flexible programming systems
and associated run-time environments that support collective
(hardware-software) optimization.

NGS sequencers such Illumina and Ion Torrent provide the option of onsite
servers or small clusters for performing the alignment and re-sequencing as a
pipeline right after the reads have been generated.  Scientific workflow systems
like Taverna~\cite{Taverna:2006}, BioKepler~\cite{biokepler:2011}, and
Galaxy~\cite{galaxy} (though Galaxy does not support distributed systems
currently) have been successful in integrating genome analysis tools for
bioinformatics. Script-based pipelines that combine executables are also often
used. Very few workflow systems have explicit support for distributed data or
dynamic data handling.  Therefore reusable approaches at extending existing
tools to support distributed and dynamic data are likely to be useful.

\subsection{ATLAS (an Example of Use of the WLCG) \label{WLCGSteve}}

The Large Hadron Collider (LHC)~\cite{lhc} at CERN in Geneva is
producing a large amount of data across a number of experiments,
including ATLAS~\cite{atlas}. The application discussed in this
section is an {\em infrastructural application} (meaning that it
consists of multiple stages that are run by different people, in this
case data acquisition, storage, distribution, and analysis). It has
the goal of allowing a physics group or a user to analyze data to
understand a specific physics channel (underlying physical process) as
recorded by ATLAS.  The set of available data grows steadily over time
as more data are collected (by the experiment) and reconstructed. This
application thus contains two stages: data storage and data
processing.

\subsubsection*{Application description}

Both physics groups and individuals run jobs on the WLCG (Worldwide LHC Computing Grid)~\cite{lhcb} as they see fit, either reading the reconstructed data or reading intermediate datasets created by themselves or their colleagues. If, in the future, resources should prove inadequate then adjustments to working practices to coordinate processing may be made to increase physics output.

The WLCG has a hierarchical architecture. CERN (the `Tier~0' site) is connected to eleven `Tier~1' (national) sites by a dedicated optical network. This may be extended in the future.  All data is stored at CERN, and as needed, various parts of the dataset are copied (and cached, i.e., replicated) to the Tier~1 sites, and from there to `Tier 2' (regional) sites and to `Tier~3' (university or group or individual) sites.

Before data is analyzed, the file to be analyzed is copied from the closest site to the SE (Storage Element) at the WLCG site where the job will run. From there it is either copied to a local disk or read from the SE.

There is no particular time constraint on the processing, although physics groups are eager to understand the ATLAS experiments and thus want to process data as quickly as possible.  In the event some data are missed during processing, the only effect is a reduction of statistical precision in the final analysis. Accurate bookkeeping is therefore important in order to know which data have been processed.

 \subsubsection*{Big data aspects}

Roughly 20~TB of data are collected every day. Multiple generations of processed data are kept. Many data files are replicated on multiple sites; this is done both by copying it to where it is likely to be useful and partly dynamically as the need arises.  The bulk of the data is individual event data at different stages of refinement. This is held in an in-house representation of C++ serialized objects. Though this format is not standard it is used by all LHC experiments.

 \subsubsection*{Distributed aspects}

The WLCG has 250,000 cores distributed over 140 sites and with 100~PB of disk. It uses a mixture of gLite software, experiment specific software and physics group specific software.  The data is distributed and replicated as described previously.

 \subsubsection*{Dynamic aspects}

The data being processed can be considered as dynamic as it grows steadily as new raw data are collected and the basic processing program (called reconstruction) is run to produce information about specific particles involved in a collision. This reconstruction is rerun periodically (two or three times per year is expected) as calibrations of the detector are improved. Reconstruction algorithms are also improved as are the physics groups' selection codes.

 \subsubsection*{Other important issues}

Data analysis is really a pleasingly parallel task, where various events are analyzed in parallel and then a summary is created of all the analyses.

The key challenge here is really in the infrastructure -- building a system that can store massive amounts of data and move them to where the processing is to be done by a large diverse group of users.

\subsection{LSST \label{astroAdam}}

This application, which similarly to the previous application is an
{\em infrastructural application}, wants to find and study variable
objects and moving objects, using a telescope that takes images each
night. This is a scenario that may be used by Large Synoptic Survey
Telescope (LSST)~\cite{lsst}, and is somewhat similar to that done by
the Palomar Transient Factory (PTF)~\cite{ptf,ptf2,ptf3} and by
Pan-STARRS~\cite{ps1,ps2}.  Overall, this application is a
leading-edge example of the work needed to build a set of data
(catalog and images) that then can be used by others for analysis (an
example of which is the next application, in \S\ref{astroSOAAdam}.)
This application contains three stages: data transformation \&
decisions, data storage.
(Data generation comes from the telescope itself,
and processing is done outside of this application, e.g. using the next
application.)

\subsubsection*{Application description}

Processing of data (data analysis and detection of new sources) from each image should be done while the next image is being taken by the telescope.  This means that if anything unusual is detected, normal observation can be interrupted. Other observing resources can also be notified instantly, so they can observe the same event. As data is collected, it is added to all the data previously detected from the same location of sky to create a very deep master image. LSST will also build up a database of all known moving objects.

Every time a new image of the sky is obtained, the master image will be subtracted from it. The result is a subtracted image that contains the difference between the current sky image and its average state.

This subtracted image is then processed by a cluster of computers with 3 main steps:
\begin{enumerate}[(i)]

\item Using the existing object catalog (containing the known orbits of all known objects), find objects which are expected to appear in subtracted image, given the area of sky and time of day.
Cross-match expected objects with sources in subtracted image, resulting in the unmatched-source catalog, which contains sources that cannot be matched with a previously known object.  If these sources can be tracked over a few images, an orbit can be determined for them and they can be added to the object catalog.  Also, the orbit catalog can be updated with re-detections of known objects; each rediscovery provides information that improves the known orbits.

\item Attempt to classify all entries in the orbit catalog (i.e., all the objects which now have known orbits). If any Near Earth Objects are detected to be passing close to the earth, alerts are generated to astronomers so that follow up observations can be scheduled.

\item If an unknown object is found that cannot be classified locally, at least two possible options exist. One is a coordinated effort of comparison with other observatories, where the local observatory issues a call for participation to a set of additional possible observatories, each of which can accept or reject the call.  The local observatory then chooses one of the accepting observatories from which to obtain data, and that observatory takes data and returns it. The local observatory then decides what to do: classify the object, contact a human, or call for more observations (automatically to other observatories that decide whether or not to accept, and when)

A second option is less coordinated, where this observatory would send VOEvent
messages, notifications of transient events which can be used to
trigger follow-up observations on other telescopes. These are small XML messages
containing basic information about the position of a transient source and a
small amount of other metadata which the operators of a given telescope can use
to define triggers for follow-up observations of interest to them or their
community.

\end{enumerate}

 \subsubsection*{Big data aspects}

LSST will generate 36~GB of data every 30 seconds, and over a 10-hour winter night, will collect up to 30~TB.  Data will be stored as FITS image, and in large databases containing analyses of  the discoveries.

 \subsubsection*{Distributed aspects}

The overall system will contain a central telescope with computing data and resources, linked to a wide area network of observatories and data and computing resources.

 \subsubsection*{Dynamic aspects}

The time constraints differ for different types of transient sources, but in some
cases notification of remote telescopes must be made within a few minutes if it is to be useful; the
overheads associated with setting up a follow-up observation mean that much
shorter time scales are practically irrelevant, although they are still
scientifically interesting. The quality constraints center on providing a
remote telescope with good enough information about a potential transient source that
it can make a reliable decision about whether to interrupt its existing
observing program to follow-up the event. In most cases, the candidate
transient will be some sort of noise event, so fairly sophisticated filtering is
required to bring the false positive rate down to something acceptable, given
the cost of observing time on cutting edge telescopes.

 \subsubsection*{Other important issues}

Data will initially be analyzed at the telescope; discoveries will be sent to other centers.

Pan-STARRS, a contemporary sky survey with similar goals as LSST,  uses a scripted programming model
on a cluster for image processing of telescope frames to extract object attributes as a CSV
file.
This is followed by followed by scientific workflows, based on Trident, that load the files
into individual databases daily, and merges them with a master distributed-database on a weekly basis on an
HPC cluster~\cite{Simmhan:Building:2009}. Data extraction and mining operations by astronomers are composed using a pseudo-SQL
language with user defined functions and submitted for execution on the master database using a batch system.

\subsection{SOA Astronomy \label{astroSOAAdam}}

An increasingly important feature of astronomical research is the
existence of systematic sky surveys covering the whole sky (or large
fractions thereof) in many different wavebands. The principal outcome
from these surveys are astronomical source catalogues, which contain
values for attributes characterizing all the celestial sources
detected in the sky survey dataset.  These are now typically
implemented in relational databases, with SQL interfaces implemented
through web pages, and the principal goal of the Virtual Observatory
(VO) being developed by the International Virtual Observatory Alliance
(IVOA)~\cite{ivoa} is to standardize access to these, and other
astronomical data resources, so that astronomers can readily perform
multi-wavelength analyses, combining all the extant data on particular
sources.  The previous application (\S\ref{astroAdam}) was an example
of a ``generator'' of such data resources, this application is a
typical ``user'' of them.  As such, it can generally be considered as
having just the last application stage: data processing.  Thus it is
an {\em infrastructural application}.  The particular example
described here calculates the photometric redshift for a given area of
sky using two tools: HyperZ and ANNz, both accessible via web services
interfaces~\cite{barker}.  However, it represents many
service-oriented architecture applications that are designed to work
with the VO.

\subsubsection*{Application description}

The VO includes a registry that contains metadata describing all the data
resources published using VO data access standards, enabling the discovery of
datasets relevant to a particular analysis. Web service implementations of data
access protocols enable users to access data in these repositories
programmatically, and distributed scratch space is provided for the storage of
intermediate result sets, so the user does not need to route large data flows
through his/her own workstation. In the future, more and more data analysis
software will be made available through web services compliant with VO
standards, enabling more of the data integration and analysis process to be
combined in workflow.

The application orchestrates a set of services through a pipeline in order to compute the redshift of a given area of sky.  It contains a set of components.  1) The Wide Field Survey Archive (WFS), which is an image catalog. 2) A spectroscopic database, used to provide spectroscopic data for an area of sky. 3) The SExtractor tool, which is an image processing that extracts all objects of interest (stars, galaxies, etc.)  4) A cross matching tool that compiles one table of all objects of interest over the five wavebands.  5) Hyperz, the first photometric redshift estimation algorithm.  6) ANNz, the second photometric redshift estimation algorithm, and 7) mySpace, the AstroGrid storage service.

 \subsubsection*{Big data aspects}

Images will be O(1) GB files.

 \subsubsection*{Distributed aspects}

The data resources published to the VO exist in a number of data centers
distributed internationally. Major datasets---e.g. those from large sky
surveys---are typically implemented in relational databases running on high-spec servers
or clusters thereof.  The services in this application are also distributed, with web service interfaces used to link them.

 \subsubsection*{Dynamic aspects}

The WFS catalog will be updated regularly, therefore a query at time $x$ will not provide the same data as a query at time $y$.  In addition, the specific queries that will be issued by users are unknown, and will change over time.

 \subsubsection*{Other important issues}

Enough data needs to be available for a given area of sky from both the WFS archive and the spectroscopic archive.

\subsection{Understand the Cosmic Microwave Background \label{astroJulian}}

This {\em traditional application} performs data simulation and analysis to understand the Cosmic Microwave Background (CMB)~\cite{cmb}, which is an image of the Universe as it was 400,000 years after the Big Bang. Tiny fluctuations in the CMB temperature and polarization encode the fundamental parameters of cosmology and, using the Big Bang as the ultimate particle accelerator, ultra-high energy physics. Extracting this information from the data gathered by current and anticipated CMB observations is an extremely computationally intensive endeavor for which massively parallel tools have been developed.  This application contains just one stage: data processing.

\subsubsection*{Application description}

The CMB community wants to extract cosmology and fundamental physics
from the observations gathered by the detectors as time-ordered
sequences of O($10^{12}$ - $10^{15}$). These observations are reduced
first to a map of O($10^6$ - $10^8$) sky pixels, then to O($10^3$ -
$10^4$) angular power spectrum coefficients, and finally to O(10)
cosmological parameters~\cite{cmb2}.

The central computing challenge for any CMB dataset is the simulation and analysis of O($10^4$) synthetic observations, used to correct for biases and quantify uncertainties in the analysis of the real data.
Preconditioned conjugate gradient techniques are used to solve for the maximum likelihood sky map given the input data (obtained by scanning the sky with hundreds to thousands of detectors for weeks to years) and its piecewise stationary noise statistics.
To avoid the I/O bottleneck inherent in the traditional simulate/write/read/map paradigm, all simulations are performed on the fly only when requested by the map-making code.

Going from the map to the angular power spectrum is the
computationally most expensive step. The exact solution scales with
the cube of the number of pixels (in the map), so going from map to
angular power spectrum is ruled out now (but was possible earlier for
much smaller observations). The approximate solution requires sets of
O($10^4$) Monte Carlo realizations of the observed sky to remove
biases and quantify uncertainties, each of which involves simulating
and mapping the time-ordered data.

The map-making application is therefore applied to both real and
simulated data, but many more times to simulated data (which requires
us to use the on-the-fly simulation module too).

 \subsubsection*{Big data aspects}

There is O(1 - 10) TB input data and a similar volume of output data.

 \subsubsection*{Distributed aspects}

This application is targeting the largest supercomputers available: Hopper, Tianhe, Blue Waters, etc. It is all about the cycles, although as the concurrency increases previously solved I/O and communication scaling bottlenecks typically re-emerge.

The application currently uses a single HPC systems, but the scientists have discussed using distributed systems, with a model of remote systems being used for the data simulations.
Each simulation would be launched from the central system that is building the map, and
output data from the simulations would be asynchronously delivered back to that central
system as files that would be incorporated in the map as they are produced.

 \subsubsection*{Dynamic aspects}

Overall, this application can make use of whatever computing it can access in order to run the simulations needed to flesh out the observed data (which can be read from local disk).  The simulations are dynamic in time and in location.

 \subsubsection*{Other important issues}

A goal is to simulate and map O($10^2$) realizations of an experiment's data in O(1) wallclock hour to provide 10\% errors during the early analysis stages, and O($10^4$) realizations in O(100) wallclock hours for the final definitive analysis.

\subsection{Sensor Network Application \label{sensorSimon}}

Marine sensing covers a number of applications, including
environmental monitoring, remote exploration, marine life surveys, and
habitat assessment. In each case, the main challenge is collecting
data reliably from a difficult working
environment, without interfering with the natural behaviors the
animals exhibit.

A canonical example is the monitoring of seal and other sea mammal
populations by the Scottish Oceans Institute (SOI), which tags animals
with sensor packages that can record dive behavior, speed, and
movement~\cite{SMSSeal,SealContact}.  These datasets are then analyzed
offline using traditional statistical techniques and integrated with
Google Earth to visualize animal tracks on a large scale.  This is an
{\em infrastructural application}. It begins after data generation,
and it includes the data storage and processing stages.

\subsubsection*{Application description}

The sensor packages on the marine mammals report back to base when the animal crawls up a beach and comes within range of a cellular network. Data are collected remotely and brought to a central site for analysis. Analysis includes statistical applications and visualization is done using Google Earth.

 \subsubsection*{Big data aspects}

The data sets collected are modest in size by scientific-data standards. They include positions and motion vectors for sea mammals, and concentrations and gradients for environmental missions.

 \subsubsection*{Distributed aspects}

Time series data is published from distributed sensors, but they are stored in a central repository. The data is analyzed locally and visualized by distributed users.

 \subsubsection*{Dynamic aspects}

Time series data is published dynamically from the sensors when they are within communication distance of cellular towers. So the time series data arrives at the central repository asynchronously relative to the time of collection.

There is some basic adaptation to the resolution of the data collected to account for different resolutions of the sensor packs.
Larger-scale environmental sensing applications may
make use of more structured and extensive adaptation, such as
changing the sampling frequency and other management characteristics in
response to the data being observed.

 \subsubsection*{Other important issues}

Collecting data reliably is challenging due to the distributed deployment of sensors in hostile environments.

\subsection{Climate \label{climateDan}}

This {\em infrastructural application} is aimed at supporting
international CMIP/IPCC~\cite{intergovernmental2007fourth,cmip}
intercomparison activity, which involves producing, storing, and
analyzing data from multiple climate simulations.  It includes data
generation, storage, and processing stages.

\subsubsection*{Application description}

The overall system is comprised of three stages.  In the first stage,
climate centers run a prescribed set of common experiments that
produce 2-10 PB of data.  Data can also come from sensors, in which
case the center would still post-process the data before it would be
published.  Centers can then publish their own output data, or send it
to another center to publish.  The data is generated over a roughly
2-year period (then post-processed and published over a few more
months).  This stage is characterized by being both
distributed-compute and distributed-data intensive.

The second stage is data storage.  The Earth System Grid Federation
(ESGF)~\cite{esgf} develops and deploys a federated network of gateways and
associated data nodes.  As models run at each center, the output data
is post-processed into packages with common formats and prescribed
metadata conventions.  Most centers deploy the ESGF data node software
stack, and using this, they manage the data from their experiments.
The data node software stack scans the data, makes sure it has the
right metadata fields, does QA/QC, and builds a set of catalogs of the
prepared data.  Minimally, the catalog provides HTTP links to data
elements, and it can also provide GridFTP endpoints, and/or product
services that abstract the dataset in other ways: get a whole file,
subset, browse, etc.  When the center is happy with the data/catalog
in the data node, they publish it to a host/affiliated gateway.  This
submits the catalog to the gateway.  The gateway then shares this
catalog with other gateways so that all gateways have a consistent
view of all the published data.  In general, most data is replicated
at several sites.  The replication activity is manually
requested/initiated by a gateway owner.  The properties of this stage
are that it is distributed data-intensive (with distributed gateways
and data nodes) and that it is dynamic (data appears in the system
over time).

The third stage is data analysis.
The approximately 20,000 users (as of Dec. 2010) can browse/search a catalog at any gateway and locate data, which might be hosted by a data node affiliated with another gateway. (This can output a `wget' script that can later fetch the data.)  They can also authenticate, and thus gain access to a group, which might have private data.  They can also download data via http, GridFTP, or access product services (the latter of which uses ESGF's data retrieval syntax---DRS---and which can be scripted.)
Some users will analyze data from a single model.  However, many applications are multi-model analyses, where many users want to look at the same parts of the output of some/all of the models, to understand if and how the models differ.
Some centers will gather some/all of the core archive (plus more, perhaps) on local systems for local users to perform ``power analyses.''
Each user's data analyses are almost always done on a single system.  Because there is no distributed computing infrastructure, this stage is search, access, and transfer intensive.

 \subsubsection*{Big data aspects}

The overall data generated and stored in the ESGF is 2-10 PB.  This includes a  core archive, which is 1-2 PB in size and contains the most popular data.

 \subsubsection*{Distributed aspects}

The data is generated by a distributed set of climate centers.  It is stored in a distributed set of federated archives.  And it is used by a distributed set of users, who either run data analyses on a climate center with which they are associated, or they gather data from the ESGF to a local system for their analyses.

 \subsubsection*{Dynamic aspects}

Data is generated over time, so the data in the ESGF changes.  One might imagine a future
version of ESGF where the launching of data analysis jobs is automated, in which case  the
location of those jobs would be dynamic, and the system might also be able to respond
(or optimize for) various types of applications.

 \subsubsection*{Other important issues}

This application/infrastructure serves a large group of users, and involves a large amount of political and technical consensus to work.

\subsection{Interactive Exploration of Environmental Data \label{envJon}}

In the environmental sciences, the use of visualization techniques is
vital for understanding the ever-increasing volume and diversity of
data that is being produced by Earth observing systems and computer
simulations.  The primary purpose of visualization is to gain insight
but the majority of scientific tools generate static plots that
neither the originating scientist nor the recipient can easily
customize to reveal new information. The issues with visualization of
a single dataset are compounded if multiple datasets are to be
examined simultaneously, as is common in environmental science for
model validation, ``ground truthing,'' quality control, and data
assimilation.

New, interactive modes of environmental data exploration and
visualization based on the principles of simplicity, open standards
and user-friendliness are required at all stages of scientific
investigation.  The application described here is an {\em
  infrastructural application} that uses stored data and consists of
one stage: data processing.  It represents a number of similar
applications~\cite{blower1,blower2}.

\subsubsection*{Application description}

The applications often take the form of graphical Geographic
Information Systems (GIS) tools, but with better support for large and multidimensional scientific data. The general vision is of a map-based interface, onto which
different datasets can be overlain.  A processing step may be initiated by rubber-banding an area of
the map and selecting from a list of algorithms that process data within the selected area, perhaps
calculating statistics of the data.  In a desktop application, this processing may take place on the user's desktop,
but in a web-based application the processing must take place on a server, which may or may not be
co-located with the datasets.  There is therefore the common problem of moving large amounts of data around in a distributed
system, to which may be added concerns of security in cases in which the data in question are not public.

 \subsubsection*{Big data aspects}

Data from instruments is small (1--100~MB) but model output may be very big (GB--TB). The data is
multi-dimensional in nature. Data aggregated over time from instruments may need to be visualized
(e.g. as an animation), causing the visualized data size to grow.

 \subsubsection*{Distributed aspects}

 Increasingly, environmental data are distributed in multiple
 locations.  The applications are driven by data served through
 distributed web services, that may in turn be ``fed'' instruments or
 computer simulation generated data. Compute services need to be
 integrated with these data services in an efficient, easy-to-use, and
 transparent way, allowing fast data processing.  The processing
 algorithms are generally simple and high overhead scheduling systems
 can cause a loss of responsiveness.  The infrastructure has more in
 common with the Web (or cloud) than the grid.

 \subsubsection*{Dynamic aspects}

Dynamic data are sometimes used, but may arise from real-time feeds from
instruments or from looking at live results from a model. Data hosted at
multiple locations may be frequently updated (often several times a day).
The data services are designed to allow server-side subsetting of large
datasets, attempting to minimize the amount of unwanted data traveling
across networks.  These make it hard to achieve scalability through caching.

In a Web environment, servers and proxy servers (which are sometimes beyond
the control of the data provider or the data user) may retain caches
of data in a strategy to reduce server load and network traffic.  For
this reason, the HTTP protocol provides mechanisms for defining expiration
times for data resources, specifying when a resource ought to be
cleared from the cache.  In the environmental and geospatial communities the Open Geospatial
Consortium specifications are being widely adopted alongside existing
protocols such as OPeNDAP~\cite{opendap}.  Unfortunately these protocols, although
built atop HTTP, do not make it easy for this versioning to be
implemented correctly.  The OPeNDAP protocol does not have the concept
of a dataset version, meaning that clients have no reliable or
efficient means to detect that a dataset has changed.  The OGC
protocols, in general, \emph{do} provide
versioning at the level of the service endpoint, but provide no
information on when a resource should be considered expired.  Clients
are therefore forced to poll the server to check for updates.

 \subsubsection*{Other important issues}

The application aims for near-real-time interaction with the data, i.e the user should not wait more than a few seconds for some kind of response to a request, so performance with low latency is a key challenge.
Security is another serious concern: many environmental datasets
are held under access control, and different providers often have very different access control policies.
In some infrastructures, these different policies are handled via a role-mapping mechanism~\cite{nerc_data}.
Concerns of access control---as well as data volume---also make it difficult for data to be replicated
to different geographic locations (and hence different access control regimes).
In a distributed environment, it is common for machines to access data
on behalf of end users (e.g., a processing service might download its input data from a remote store).
Finding a means for the user to delegate his/her authority to a multi-web-service infrastructure is
a key current challenge.  The MashMyData project~\cite{mashmydata} is investigating two solutions, based on Grid proxy certificates and OAuth.

\subsection{Power Grids \label{power}}

Demand response (DR) optimization~\cite{ferc2011DR} in smart power
grids deals with ensuring an adequate supply of electricity by
targeted curtailment of power load by consumers during periods of peak
load to avoid blackouts. One application used for DR optimization is
power demand forecasting at coarse and fine temporal and spatial
granularities to allow load curtailment operations to be triggered in
advance of a demand-supply mismatch situation.  This is a {\em
  infrastructural application}.  It begins after data generation, and
includes the data transformation \& decisions, storage, and processing
stages.

\subsubsection*{Application description}

Power usage information is aggregated from smart meters at consumer premises into the utility's data center~\cite{Simmhan:buildsys:2011,simmhan2011TR}. Data is collected at approximately 15-min intervals and transmitted to the utility through cellular and wireless networks using proprietary protocols. Complex event pattern matching over the power usage events and other information source such as weather and scheduling data accessed using web services help perform near term load forecasting. Long term power forecasting at the city-scale or at a smaller micro-grid scale (e.g. industry complex, university campus) is done using power usage models built using machine learning algorithms running at the utility's or micro-grid's datacenter using historical usage data~\cite{Aman:dddm:2011}. These applications run on private clouds and are composed using DAG, Complex Event Processing and MapReduce structures~\cite{zhou:2012:scepter}.

 \subsubsection*{Big data aspects}

Constructing forecast models using machine learning requires access to historical data that can grow to large sizes~\cite{Yin:mapreduce:2012}. Several hundred TB per year of smart meter power usage data  can accumulate for a large city with millions of customers. Other data used in predictive modeling, such as event/people schedules and building/facility features, are smaller, on the order of GBs per year. In future, social network data from consumers may also be mined to learn about power usage patterns. These can also be on the order of GBs.

Real-time forecasting uses smaller data---events on the order of 1~KB, but potentially from millions of sources---that are passed to complex event pipelines or predictive models that run continuously. These pipelines can process GBs of data per hour.

 \subsubsection*{Distributed aspects}

Data from domains such as power systems, building management systems, weather and
traffic, and social networks need to be integrated. While the
data are generated and/or stored in distributed locations, they are typically aggregated at the utility's datacenter for analysis. Limited processing may happen locally at the smart meters or local micro-grids with the results pushed as events to the utility.

The applications themselves run on private cloud platforms across hundreds of VMs. Communication between the utility and smart meters is bidirectional, allowing control and pricing signals to be sent from the datacenter to the smart meters. Integrated data will also need to be shared with external applications running on, say, mobile platforms or portals, and with regulatory agencies.

 \subsubsection*{Dynamic aspects}

The data rates for events published by the sensors may change, either passively due to local measurement limitations or actively throttled by the utility to meet the application's accuracy needs/resource constraints~\cite{Simmhan:sciencecloud:2011}. The data from smart meters can be sent synchronously as they are generated or in batch mode once a day to conserve bandwidth. The application processing itself is done elastic resources, scaling up or down to meet latency requirements. Some of the forecasting models can also decide to listen to additional data streams on-demand when higher accuracy is required or special situations are encountered.

 \subsubsection*{Other important issues}

Semantic
information integration is being performed using an ontology composed together from individual
domains~\cite{Zhou:itng:2012}. Machine learning models using Matlab, Weka and Hadoop are being used. Complex event processing is using the ``Siddhi'' CEP engine with semantic support added~\cite{siddhi}. Stream processing on Eucalyptus using IBM Infosphere Streams is being performed.

\subsection{Fusion (ITER) \label{fusionScott}}

With ITER~\cite{iter} scheduled to operate perhaps around 2022, plasma fusion has become
one of the highest priority research items for the US DOE. Scalability of
first-principles fusion codes is an important part of fusion research
needed to obtain predictive capability of plasma performance with
first-principles physics fidelity. The goal is to build a realistic
full-distribution kinetic code, in realistic ITER geometry, combining
the strengths of each of the existing gyrokinetic codes (GEM, GTS,
XGC, and GTC), and utilizing the most up-to-date computer science,
hardware design, and applied mathematics technologies developed. This
{\em traditional application} is intended to be capable of 1) discovering
macroscopic magnetohydrodynamics activity in burning plasmas, 2)
understanding Energetic particle effects in fusion reactors, 3)
understanding the heating in a fusion reactor, and 4) understanding the
interaction of the plasma at the material boundary. This applications just
has the data processing stage, along with associated data movement.

\subsubsection*{Application description}

The applications that are currently run (XGC1, XGCP, GTC, GTS) run on all of the DOE and TeraGrid leadership class facilities, and require very large resources for the modest problems that are currently being solved. In the next generation (exascale), not only will all of the computing power that will be available at the time be needed to understand ITER, all of the data that will be generated during the simulation also will have to be understood. The community's techniques have been to examine and understand the data in situ, to reduce the total amount of data produced in the simulation.

 \subsubsection*{Big data aspects}

Today the team has produced O(100) TB per week for a simulation using the ADIOS middleware~\cite{adios} incorporated in the simulations. Using staging techniques, they believe that they will only produce O(10) PB on an exascale machine.

 \subsubsection*{Distributed aspects}

The team always runs on all available resources, so if they have computer time
across the country, then their data will be distributed, and so will the
computation.

 \subsubsection*{Dynamic aspects}

The primary aspect that is dynamic is that the set of computing resources that are used
for a given run are based on those that are available at the time of that run.  As part of this,
data is streamed from one running simulation to another, and is transformed in-flight.

 \subsubsection*{Other important issues}

The applications run on a distributed set of leadership-class
facilities, using advance reservations to co-schedule the simulations.
Each code reads and writes data files, using ADIOS and HDF5.  Files
output by each code are transformed and transferred to be used as
inputs by other codes, linking the codes into a single coupled
simulation.  Because the overall data generated is so large, it can
not all be written to disk for post-run analysis, so in-situ analysis
and visualization tools are being developed.

\subsection{Industrial Incident Notification and Response \label{Kees}}

Applications such as large scale crisis management, maritime security, disease and pollution
monitoring, and environmental control need to fuse large amounts of heterogeneous data to form virtual
information sources that mitigate noise~\cite{emergency1,emergency2,emergency3,emergency4}.  The area of greater Rotterdam in the Netherlands is a
chemical hub and houses more than 15\% of the Dutch population, making safety and security a major
concern.  Some leakage of chemicals into the air is an everyday, unavoidable event.

The environmental protection agency (EVA) maintains a command and control call center staffed by
chemical hazard experts to (1) monitor the air quality and sewage, (2) in case
of an incident locate the source, (3) identify the threat and (4) provide the public authorities, the
emergency response organization, and the public with situation awareness and suggested
actions.  The information that is shared is often incomplete, uncertain and
inconclusive. Knowledge within the EVA as well as from outside experts and industries must be accessed to meaningfully use that information.  This subsection describes the {\em traditional application} that the EVA uses for these tasks.  The application starts after data generation, and includes data transformation \& decisions, storage, and processing.

\subsubsection*{Application description}

For monitoring and incident detection, continuous streams of data are processed and scanned for
anomalies from a small number of physical sensors and air analysis data (gas-chromatographs
outputs).  If an incident or an anomaly is detected, new streams of data become available from the general public and field inspectors and must be processed. Static information on weather
and about chemicals are also needed.  These may be stored in
different places, access may not be granted automatically, and validity and update frequencies may not
match expectations. Thus specialized pre-processing is needed.

Different knowledge-based processing models must be used account for uncertainty in
the data and sensor behavior, and provide robust fusion (processing) for situation awareness and
decision support. Some confidential data must be processed \emph{in situ}. New
intermediate and final data that are generated may have different retention characteristics and
need to be stored and made accessible for visualization and analysis in different workflows.

Processing in done in semi-real-time or in batch mode processing on a
small, private computer infrastructure. Therefore, scalability is
lacking and the dynamics of the workflows is kept under strict human
control to match the capabilities of a set of static resources.

 \subsubsection*{Big data aspects}

The data sizes are not very large.

 \subsubsection*{Distributed aspects}

A number of special data and sensor interface processes will be distributed over various platforms
and fixed in place.  A low frequency stream of reasonable fidelity data from a small number of
physical sensors and an even lower frequency stream of air analysis data (gas-chromatographs
outputs) is continuously trickling in.  Low fidelity human provided data from local people complaining to the call center,
field inspectors calling in with observations and measurements, information from
social media, pictures from cell phones and emails sent to a centralized email account are entered manually into a local databases.  Static information such as current
weather conditions and forecast and information about chemicals are also needed.  These data are
made available via a distributed shared data space.  Some confidential data must be processed remotely and only aggregated results communicated back to other entities. The information processing algorithms are packaged as autonomous software agents that are managed by a multi agent system infrastructure
(middleware) and humans (via GUIs). Processing algorithms need to access multiple concurrently and remote data stores.

 \subsubsection*{Dynamic aspects}

Both the data and the processing are dynamic in nature. Data that cannot be moved due to
confidentiality have to be processed remotely by moving applications to them.  A process integration
framework will use agents that provide autonomous and composable resources to dynamically construct
workflows for data driven processing.  If an incident happens or an anomaly is detected, it triggers
new streams of data about other anomalies that must be processed.  Another source of dynamic
behavior is when the event is escalated and additional management processes come into play, each
requesting and adding their own sort of processing to the mix.

 \subsubsection*{Other important issues}

Quality of service aspects that are important for real world and real-time
applications are the dynamics of scheduling from an application perspective,
multi independent level of security support, processing service discovery support, autonomous reconfiguration support and logging-for-traceability.
Most of the dynamic data that these applications use has a short validity
period, i.e., it must be processed now or we process the next update coming
in. The data is uncertain, i.e., it may have dynamically changing signal to noise ratio
and although expected at some base frequency, may not be available for some
period of time.

\subsection{MODIS Data Processing \label{modisKeith}}

This {\em traditional application} allows the coordinated usage of MODIS~\cite{modis}
satellite imagery, and its integration with ground based sensors and models for
environmental applications.  It consists of data storage and processing stages.

\subsubsection*{Application description}

The application~\cite{modisAzure, modis:ipdps:2010} is used to handle the processes of downloading data from different
sites and integrating them. The integration consists of re-projecting the data from one map projection to another,
spatial and temporal re-sampling, and gap filling. The application itself is a loosely-coupled data
pipeline that coordinates these processes.

In addition, scientists can submit Matlab scripts for further custom processing on subsets of
the data that has been integrated. This includes some summarization processing and plotting
routines.  The only time constraints are on the debugging cycle for debugging the scientist-supplied
reduction codes. These are purely driven by the users' patience.

The application has been implemented in both the Windows Azure cloud
environment~\cite{win_azure} and a standard HPC environment.

 \subsubsection*{Big data aspects}

There is approximately 25~TB of data for one global year.  The main data is the various MODIS satellite data
products, of which 11 are used by this application. These are in approximately 3,000,000 files for one global year. A
smaller set of sensor data from the fluxnet data network is used as input to the evapotransporation
calculation, along with a set of 20-30 files containing various metatdata like climate regions used
by the calculation.

 \subsubsection*{Distributed aspects}

 The data is collected from a wide variety of FTP servers managed by
 different groups and brought into the cloud/HPC environment. Custom
 software tooling manages and validates this download
 process. Reliable cloud message queues are used to manage the
 coordination between the different components in the multi-stage
 pipeline. Simple NoSQL tabular storage is used for logging and fault
 recovery. Data is stored in cloud blobs and NoSQL table
 storage. Application scripts on the user's desktop can be uploaded
 into the cloud as post-processing routines and analyses on the
 integrated data.

 \subsubsection*{Dynamic aspects}

 The data is currently static and the application acquires the data
 from a set of FTP servers on a schedule or trigger. But as the
 pipeline is expanded to handle other data, the usage of real-time
 streaming sensor data is also expected. There is infrastructure
 dynamism through the use of cloud virtual machines, acquired and
 released on-demand, to execute tasks in the pipeline and to execute
 the user's application scripts. While the pipeline structure itself
 is static, there is a limited form of application dynamism through
 the use of user specified scripts pushed into the cloud for execution
 over the integrated data.

 \subsubsection*{Other important issues}

The data is in one of two map projections, and multiple temporal and spatial resolutions. The
fluxnet sensor data that is used required significant data integration, but was handled in a
separate project.

The quality constraints are driven by the needs of the scientific reduction application. Significant
effort is required to ensure that the data handled to the scientific application meets these
requirements.

\subsection{Distributed Network Intrusion Detection \label{detectionJon}}

This {\em traditional application} performs detection of distributed network intrusion and distributed denial of service attacks~\cite{jon}.  Network TCP/UDP
traffic data is analyzed/mined at each site. Potential alerts are collected centrally and further
analysis is done to see if an attack may be underway by correlating alerts, and events are then
communicated back to local sites. The following features are extracted: 1) basic features such as
source/destination IP address pair, port number, and protocols and 2) some additional features
such as the number of flows from the same source to specific destination IP addresses for a
predetermined time period. The extracted features are input for the MINDS system. While well-known
intrusions are detected by the known attack detection module, the anomaly detection module is
applied to the remaining network connections. During the process of anomaly detection, a training
set is generated if not given as input. The anomaly detection module scores each network
connection to reflect the level of anomalousness of the network connection compared to the normal
network traffic. The highly scored network connections are analyzed by the association pattern
analysis module to produce summary and characterization of the possible attacks. This summary
and characterization is used to create new signatures and models for unknown emerging attacks.
MINDS contains various modules for collecting and analyzing massive amounts of network traffic.
Typical analyses include behavioral anomaly detection, summarization, scan detection and
profiling.  The application includes all four stages: data generation, transformation \& decisions, storage, and processing.

\subsubsection*{Application description}

MINDS is deployed at each site as a grid service. The analysis application continuously scans local logs, and reports interesting events to a centralized analyzer, and alerts are reported back to local sites.

Each site uses a data capturing device such as network flow tools or network monitoring tools like tcpdump  to collect the network traffic data from routers.  This collected data is first filtered to remove unimportant network connections due to the large data volume.

All communication in the system is accomplished using messages on top of HTTP.

 \subsubsection*{Big data aspects}

For local data, this is site dependent. It can be MB/hour/site.

 \subsubsection*{Distributed aspects}

The system is deployed across clusters
located  at a set of participating sites (University of Minnesota, University of Florida, and University of Illinois, Chicago) connected by the public Internet. In addition, the data generated is also produced (distributed) at this set of sites.

 \subsubsection*{Dynamic aspects}

Network traffic data is highly dynamic and constantly streaming in to the application. The data consists of TCP flow and UDP records, which are very structured.  They contain IP addresses, ports, various headers, and payloads. Interesting events are propagated across sites. These are very small, a few KB at most.

 \subsubsection*{Other important issues}

 The application is built using Globus Grid Services for wrapping the
 MINDS analysis applications and for providing storage services for
 events. At the time the application was written, the project had to
 develop tools such as remote launching of Grid services
 themselves. MINDS was also a separate software package that was
 utilized. The project also built an authorization service for
 controlling access to data products. Other noteworthy points:
 real-time is ultimate goal of the project and false positives are
 more tolerable than false negatives.  Finally, the project found it
 to be very difficult to obtain network data at each site due to
 privacy concerns.

 IDS systems also use pattern detection tools such as Bro and Snort,
 which perform simple string and regular expression pattern matching on
 IP packet headers and body. These are monolithic tools that run
 standalone, without any comprehensive programming support to automate
 updating the patterns or taking action upon pattern detection.

\subsection{Application Summaries}

Here we provide short summaries of the computational (as
opposed to scientific) aspects of each
application in preparation for the analysis that
follows in the rest of this document, which is motivated by and
centered around these set of applications.

{\bf NGS Analytics (\S\ref{bioSilvia})}: Data is processed through a
customized pipeline of analysis tools, such as BWA, BFAST, Bowtie;
processing of each data element is independent of other elements, so
processing is decomposed/parallelized over the data elements.  All
data is stored in files.  Later analysis may want to operate on files
from multiple data stores distributed around the world, e.g., to
compare O(1000s) of TB-scale datasets (metagenomics) with each other.
Dynamic elements include the choice of executables in each instance of
the pipeline, the design of the pipeline itself, and the choice of
computing infrastructure to provide the best resource usage and
turnaround time.  The solution to this problem is not clear.  There is
a need for flexible programming systems and associated runtime
environments for hardware-software co-optimization.

{\bf ATLAS/WLCG (\S\ref{WLCGSteve})}: There is a hierarchy of systems. Data are centrally stored, and locally cached (and copied to where they likely will be used), perhaps at various levels of the hierarchy. Processing is done by applications that are independent of each other. Processing of one data file is independent of processing of another file, but groups of processing results are collected to obtain statistical outputs about the data.

{\bf LSST (\S\ref{astroAdam})}: Data is taken by a telescope. Quick analysis is done at the telescope site for interesting (urgent) events (which may involve comparing new data with previous data).
The system can get more data from other observatories if needed, request other observatories to take more data, or call a human. Data (in files) is then transferred to an archive site, which may be at the observatory. At the archive site, the data are analyzed, reduced, and classified, some of which may be farmed out to grid resources.  Detailed analysis of new data vs. archived data is performed.  Reanalysis of all data is done periodically.  Data are stored in files and databases.

{\bf SOA Astronomy (\S\ref{astroSOAAdam})}: Services are orchestrated through a pipeline, including a data retrieval service that is used to share data across VO sites, as well as analysis and image processing services.  Data (files) are moved through the pipeline, and intermediate and final products can be stored in AstroGrid storage service, such as mySpace.  Scratch space is provided for storage of intermediate results.

{\bf Cosmic Microwave Background (\S\ref{astroJulian})}: This application builds a sky map.  It reads data from detectors, and obtains data from simulations that are performed on-the-fly, when requested by the map-making component (in a master-worker sense), one of a set of components that gradually reduces the data from the detected and simulated data to a small number of cosmological parameters.

{\bf Sensor Network Application (\S\ref{sensorSimon})}: Data are collected from mobile sensors that periodically transmit when they come within communication distance of receivers. Fault tolerance (collecting data in a hostile environment) is an important component. The data, which are very diverse, because they can involved different resolutions, sampling frequencies and other management characteristics., are brought to a central site. Stored data are analyzed using statistical techniques, then visualized with tools such as Google Earth.

{\bf Climate (\S\ref{climateDan})}: Climate simulations are run by climate centers.  Data (outputs of climate simulations) are distributed in multiple federated stores.  Current data analysis and visualization tools require the user to bring the data they want to analyze to a local system, then run the tools locally.  New tools will include the capability to add data transfer.  This will allow the set of tools to iterate: transfer data, do analysis, etc.

{\bf Interactive Exploration of Environmental Data (\S\ref{envJon})}: Applications are driven from map-based graphical tools.  The inputs are geographic areas and algorithms to run on the data in those areas. Data can come through web services from stored data, sensors, or simulations (possibly running).  The services can perform server-side processing to reduce the size of data files to be transferred. Responses are provided in near-real-time.  Security also supported through a delegation-based model.

{\bf Power Grids (\S\ref{power})}: Diverse streams arrive at a central utility private cloud over cellular and wireless networks at dynamic rates controlled by the application. Communication is bi-directional, with pricing and control signals sent to customers power meters.  A real-time event detection pipeline, involving complex event processing and machine learning/data analysis algorithms and using other stored information such as weather data, can trigger load curtailment operations. Data mining is performed on current and historical data for forecasting. Partial application execution on remote micro-grid sites is possible.

{\bf Fusion (\S\ref{fusionScott})}: Multiple computational intensive
physics simulation codes run concurrently on a distributed set of
parallel computers.  Data from some codes are streamed to other codes
to link them into a single simulation.  Data are transformed in flight
into needed inputs.  Data are also examined in situ to understand the
overall state of the simulation.  Some data are stored for later data
mining and visualization.

{\bf Industrial Incident Notification and Response (\S\ref{Kees})}:
Data are regularly streamed from diverse sources, and sometimes
manually entered into the system. Initial disaster detection causes
additional information sources to be requested from that region and
one or more applications to be composed based on available data. Some
applications run on remote sites (where the data is stored) for data
privacy. Escalation can cause more humans in the loop and additional
operations.

{\bf MODIS Data Processing (\S\ref{modisKeith})}: Data are brought
into system from various FTP servers.  A pipeline of initial
standardized processing steps on data is done on clouds or HPC
resources.  Scientists can then submit executables and Matlab scripts
that do further custom processing on subsets of the data, which likely
include some summarization processing (building graphs).  The
application has been run on a cloud and in an HPC environment.

{\bf Distributed Network Intrusion Detection (\S\ref{detectionJon})}:
Data are analyzed or mined by a service running at a number of local
sites. The results are sent (via HTTP messages) to a central site for
analysis. Events are then communicated (via HTTP messages) back to
local sites.

\section{Understanding Distributed Dynamic Data}\label{sec:distdyndata}

In the previous section we described thirteen applications and
analyzed them along an analytical space
spanned by the three ``D''s of data-intensive, distributed, and
dynamic behavior.

\emph{Distribution} refers to the presence of application data in
different physical or logical locations. Distributed data may arise
naturally in that the data sources themselves are geographically
separated, such as from remotely located sensors.  In some cases, data
must remain distributed for privacy and policy reasons, or due to
constraints in sharing.  In other cases, a full dataset may not fit in
one location either on disk or in memory and must be decomposed to
enable storage and/or computation. Distributed data may be driven by a
need for performance, scalability localization, reliability, or
availability considerations. For example, copies of data may be
generated and distributed to maintain data availability or load
balancing.  In general, a \emph{distributed application} is one that
needs, or would benefit from the use of, multiple
resource~\cite{dpa_surveypaper}. Example benefits include increased
throughput, decreased time to solution, and increased reliability.  A
distributed application rarely consist of a single component that is
distributed; most often a distributed application is composed of
smaller and functionally self-contained, if not entirely independent,
components that can execute on different distributed resources.

\emph{Dynamism} refers to changes in the behavior and characteristics
of an entity %
over time and/or space.  Over a long enough timescale, almost all
infrastructure, and especially distributed infrastructure, will
exhibit some form of dynamism, as a consequence of failure, varying
load, etc. However, we are interested in dynamism on timescales that
influence application behavior, scheduling decisions, etc. Some
prominent examples are resource fluctuation, data availability, and
change in application execution characteristics.

In this section we will try to extract commonalities in the dynamic
behavior of applications studied in the previous section, as well as
properties and types of distribution seen in these applications.

\subsection{Types of Distribution \label{sec:distTypes}}

There are many scenarios under which distribution of data is either
necessary or desired. In addition to the reasons behind distribution
presented at the beginning of this section (viz., large volumes of
data) and reasons that constrain or localize data (viz.,
privacy/policy issues), another common situation under which data is
distributed occurs when data comes from multiple sources, for example
multiple sensor streams or data centers, but needs to be processed
collectively.

There are many ways to address distributed data.  These include but
are not limited to:
\begin{itemize}
\item \textbf{Replication:} Data is duplicated on multiple nodes. When
  fully replicated, each execution unit has access to the same
  data. Replication can be partial, i.e., data can be replicated to a
  subset of resources. The degree of replication quantifies the
  average of the number of copies of each data item. Replication may
  be non-uniform, e.g., if only a subset of data (perhaps ``hot''
  data) is replicated. Caching is a lightweight form of replication.
\item \textbf{Partitioning:} Data is divided and distributed across
  multiple nodes. Different forms of partitioning exist.  For example,
  data can be thematically partitioned with respect to its type, such
  as sensor data vs. simulated data, or data can be partitioned
  according to a spatial or temporal attributes. This can be an opaque
  (e.g., ID) and/or a domain dependent (e.g., spatial) attribute.
\item \textbf{Streaming}: A form of distributed data processing in
  which data arrives from a source; often the stream is a steady,
  continuous flow, but often it fluctuates both spatially and
  temporally. Also, typically, the stream of data must be processed in
  near real-time. 
\end{itemize}

Which of the above approaches is employed cannot be reduced to a simple
rule. Which option can or should be used often depends upon a plethora
of co-dependent factors, such as the data volumes, the {\it degree of
distribution} (which describes the number of data sources and sites used for
storing and processing the data), the availability of compute resources, or the
desired level of reliability. Data distribution can be difficult to manage for
reasons of performance: latencies and bandwidths may vary unpredictably.
Further, issues associated with reliability may also need to considered. An
example is the occurrence of a partial failure, where some (but not all) of
the required data remains available, leads to applications having to decide
whether to continue with reduced data or exit without processing what remains
available.

Viewed from an application's perspective, there are at least two
application characteristics that influence the distribution of
data. The first is the stage and the second is the degree of coupling
between components of a distributed application. Distribution in the
first stage is primarily determined by the structure and layout of the
data sources. For example, the placement of the data generated in the
climate application (\S\ref{climateDan}) is based on which site was
assigned to run the particular model that generates the data. But
analysis of the data from multiple models requires bringing the data
together to a single site.  Although it is difficult to formalize the
difference due to the large fluctuations between different
applications, but in general for a given application, there are
differences between the distribution prior to the application's
processing stage and the distribution of data within the processing
stage, e.g., to aid data-parallel processing.

The second application factor that determines distribution of data is
the type of separation and coupling of the components of the
distributed application. Different coupling types place different
constraints on data distribution, either due to latency, or due to
data volumes that must be coordinated. %
\emph{Loosely-coupled} distribution involves coordination and
cooperation between largely independent application components, as
might be found in a workflow system. \emph{Tightly-coupled}
applications involve greater constraints between the components,
either in latency tolerance, coordination requirements or relative
scheduling. \emph{Decoupled} distribution results in the separation of
application components elements of the same task, for example, sharing
a spatial dataset between components that cooperate to perform a
single logical task~\cite{dpa_surveypaper}.
Most of the applications in \S\ref{sec:scenarios} are loosely-coupled,
with the CMB (\S\ref{astroJulian}) and Fusion (\S\ref{fusionScott})
applications as examples of tight coupling.

\begin{table}[h]
  \begin{scriptsize}
    \begin{center}
      \caption{Distributed Processing}\label{tab:app_distributed_processing}
      \begin{tabular}{|p{3.5cm}|p{10.2cm}|}
		\hline
		\textbf{Application}  &\textbf{Distributed Characteristics} \\		
		\hline Next Generation Sequencing (NGS) Analytics &
                Data sources (NGS machine storage) and compute not
                always co-located; multiple data sources in case of
                metagenomics
		\\
		\hline ATLAS (an example of use of the WLCG) &Source
                data is generated by one central data source. Data
                needs to be distributed to match distribution of
                computing resources. During processing data is copied
                from closest replica to the resource where
		compute job is run. \\
		\hline LSST &Multiple different kinds of data sources
                (observatories, data storage); multi-step workflow
                that processes and archives data,
		carried on distributed compute resources.\\
		\hline SOA Astronomy Applications &Data resides in
                multiple data centers; services are placed closed to
		the data.\\
		\hline Cosmic Microwave Background &Replication of
                subset of data
		to compute resources for processing.\\
		\hline Sensor Network Application & Data is moved from
                different data sources to a central analysis
		system for processing\\
		\hline Climate (ESGF)
                &A subset of the data is commonly moved to compute (desktop or grid resource)\\
		\hline Interactive Exploration of Environmental Data
		&Data is brought to grid resource or desktop \\
		\hline PowerGrids
		&Data is brought to central private cloud for analysis\\
		\hline Fusion (ITER) &Simulations run wherever
                resources are available; in-situ analysis of large
                output data; partitioning/replication of data for
		data-parallel processing\\
		\hline Industrial Incident Notification and Response
                &Distributed data
                sources. Processing takes place on multiple distributed resources. \\
		\hline MODIS Data Processing & Data is copied
                (replicated) from different sources for processing.
                Processing can take place in different locations,
                e.\,g.\ cloud and
		user desktop. \\
		\hline
                Distributed Network Intrusion Detection  &Distributed data and compute resources.\\
		\hline
     \end{tabular}
    \end{center}
  \end{scriptsize}
\end{table}

\subsection{Types of Dynamism \label{sec:dynTypes}}

\begin{table}[h]
  \begin{scriptsize}
    \begin{center}
      \caption{Understanding Dynamic Characteristics. This table
        highlights different types and scenarios for dynamic data in
        the application set investigated %
          }
          \label{tab:app_dyn_char}
      \begin{tabular}{|p{4cm}|p{9.5cm}|}
		\hline
		\textbf{Application} &\textbf{Dynamic Characteristics} \\		
		\hline		
		Next Generation Sequencing (NGS) Analytics
		& Data Dynamism: different data formats and sources need to be processed\\
		\hline
                ATLAS (an Example of Use of the WLCG)
		& Data Dynamism: new raw data is steadily collected\\
		\hline
                LSST & Infrastructural Dynamism: Decision making based on observed data\\
		\hline
                SOA Astronomy Applications
		&  Application and Data Dynamism: data is continuously updated, dynamic queries\\
		\hline
                Cosmic Microwave Background Mapping
		& Application Dynamism: dynamic utilization of resources\\
		\hline
                Sensor Network Application
                & Data Dynamism: rate and volumes of data dynamic \\
		\hline
                Climate (ESGF) & Data \&Infrastructural Dynamism: new data is generated over time. Dynamic distributed infrastructure for storing and processing of data
                \\
		\hline
                Interactive Exploration of Environmental Data
		& Infrastructural and Data Dynamism: Various data sources, varying
		input rates and different queries.\\
		\hline
                PowerGrids
		& Application, Infrastructural and Data Dynamism: data itself changes,
		data rates may change (throttling), data sources change (traffic,
		weather, social networking), data queries respond to seen data (e.g.
		accuracy adjustment) \\
		\hline
                Fusion (ITER)
                & Application and Infrastructural Dynamism: data streamed
				between simulations (multiple distinct data flows)  \\
		\hline
                Industrial Incident Notification and Response & Application
				and Data Dynamism: data itself and data rate changes, dynamic
				data sources, processing changes with data\\ %
		\hline
                MODIS Data Processing
		& Application and Data Dynamism: data that is operated upon changes,
		infrastructure config changes\\
		\hline
                Distributed Network Intrusion Detection &
                Infrastructural, Application and Data Dynamism: data highly
				dynamically changing, data rate variable,
				application responds to data, on-the-flight processing\\
		\hline
     \end{tabular}
    \end{center}
  \end{scriptsize}
\end{table}

There are often spatio-temporal variability in the values and
volumes~\cite{deRoos2011UnderstandingBigData} of data.  The
applications in \S\ref{sec:scenarios} have variability in the
production/generation rate of data, and in the configuration of data
production/generation.  Data dynamism may arise from extrinsic
factors, for example because sensors begin to observe more events in
the same data stream. Contrariwise, data dynamism may occur because an
application component requested an increased sampling frequency.
Similarly application dynamism may occur through managed replication
of a compute-intensive component or from the demands of a particularly
complex query. And infrastructure dynamism may arise because the
infrastructure expands to meet additional requirements, or contracts
because of device failure.  A structured analysis of the applications
in \S\ref{sec:scenarios} suggests that the sources of dynamism can be
classified into three categories.
\begin{enumerate}

\item \emph{Data} dynamism arises when some property of the input data
  or its delivery changes, for example in terms of arrival rate,
  provenance, burstiness, or source.  There may be variability in the
  structure of the data, for example, data schema, file formats,
  ontologies, etc.

\item  %
  \emph{Application} dynamism
  involves changing the processes or components applied to a dataset,
  and might encapsulate changes in workflow, \textit{ad hoc} queries,
  or parameterization.  There may be variability in the data queries
  enacted by the application, or in the behavior
  and/or characteristics of the
  application.  %

\item \emph{Infrastructure} dynamism occurs when the system wants to,
  or is forced to, change its demands on its underlying platform, for
  example through elastic computation or partial failure.  Changes in
  the structure and load of the infrastructure can be associated with data.
  Examples include variability in the producer/provenance of the data,
  in the performance, and in the quality of service of the
  infrastructure.

\end{enumerate}

Based upon this classification, in \tablename~\ref{tab:app_dyn_char}, we
describe the applications discussed in \S\ref{sec:scenarios}.  Many of
these dynamic aspects are related: an increase in the rate of data
generation may lead an application to increase its degree of
parallelism to maintain throughput, which may in turn trigger the
allocation of new compute nodes. (In the other, less attractive
direction, decreasing computational capability might cause a decrease
in parallelism and a consequent need to drop data. Avoiding these
cases means treating dynamism equally in both directions.)

This should not be construed to imply that every data-intensive
application has all elements of data and compute dynamism and is
always distributed, but is a reflection that at extreme scales,
dynamism and distribution become increasingly important issues that
are often correlated.  As alluded to, a high degree of distribution
often results in a high amount of infrastructure dynamism,
caused by, for example, resource fluctuations and a higher failure
probability.  Similarly, high levels of dynamism, for example, in the
data sources, data rates, and queries, usually correlate with a high
degree of distribution.  For example, the Interactive Exploration of
Environmental Data application (\S\ref{envJon}) manages various
sources of data, from archived datasets to real-time feeds.
Depending on the spatiotemporal properties of data, computation can be
carried out on a set of distributed grid resources or the user's
desktop, i.e., distributed processing is used to efficiently handle
dynamic data.

\subsection{Distributed Dynamic Data Observations}

Reviewing \tablename{s}~\ref{tab:app_distributed_processing} and \ref{tab:app_dyn_char},
which have summarized the distributed and dynamic properties of the D3
applications, we observe:

\begin{itemize}
\item Data generation and consumption are commonly distributed from
  each other if the volume permits.  The Fusion application is a counterexample,
  due to data volumes and velocity.

\item Details of the storage system are often not known. Often, data is
  stored in a distributed way, e.g., on a distributed file system in the same
  data center (locally distributed).

\item Aggregation of data can either be the equivalent of gather or
reduce (using the language of collective communication). Gather is
in many ways the reverse process of distribution (called broadcast in the
language of collective communication). Similar to distribution, aggregation is a
commonly occurring pattern for data analysis.

\end{itemize}

\section{Infrastructure}\label{sec:infrastructure}

We define {\em infrastructure} as the hardware and software that is
provided to support an application, rather than being explicitly
created or hosted by the user themselves. In many cases the precise boundary between
infrastructure and application is blurred, e.g., a catalog is
a component with functionality that may be similar to that of similar
components in other applications, but is often implemented
specifically by an application rather than being provided as part of
the infrastructure. This may also be the case when reconfiguring an
infrastructure service for an application, when it is more onerous than
reimplementing the service specifically for that
application. We also introduce the concept of an ``infrastructure'' application as a collection
of infrastructure components that are packaged so that they can be used by many
applications in defined ways. The difference between these and ``traditional'' applications
is often delineated by who owns and administers the application, with traditional applications
``owned'' entirely by the user, and infrastructure applications owned by someone other than the
user but provided for use by others. Infrastructure can be specialist as opposed to generic, for instance
the purpose-built infrastructure in ESGF (\S\ref{climateDan}) that is provided to support the climate community.

In this section, we focus on the {\em software} infrastructure that is
provided to support D3 applications, specifically components that deal
with distributed or dynamic data. The components that make up the infrastructure can be categorized
based on the function provided, namely (i) infrastructure that supports {\it data management},
(ii) infrastructure that supports {\it data analysis}, and (iii) infrastructure that supports the structure of the applications.
Each of these components may be present at three different levels of the infrastructure: programming frameworks, services and tools (e.g., MapReduce, workflow engines); middleware and platform services (e.g., databases, message queues); and system-level
software capabilities (e.g., notifications, file system consistency).

\subsection{Infrastructure to Support Data Management}
\label{sec:infra_mgt}

{\em Data Sources} are used in the acquisition stage as inputs of data
into the application. Broadly speaking these sources include types such as
files, servers, sensors, instruments, simulations and models.
Additionally, databases are often considered as an intermediate data
source in some systems. The dynamic characteristics are primarily
due to availability. These are often accessed in the Data Generation stage of the application
(\figurename~\ref{fig:figures_application-stages}).

{\em Data Storage} components are used to provide persistent access to
data by the application. They may be structured (e.g., databases,
No-SQL tables) or unstructured (e.g., collections of audio
recordings). The dynamic characteristics are primarily due to the
level of persistence, e.g., temporary, durable, permanent (archived or
published), or replicated; and also in the structure of the data
stored, e.g., schema free tables as opposed to relational tables. There
may also be dynamic characteristics of storage due to rate of growth
of data, e.g., a field survey may bring in a lot of data during summer
but none during other times. Some storage components like files and databases may also serve as data
sources. These are commonly used in the Data Storage stage of the application (\figurename~\ref{fig:figures_application-stages}),
but may also be present at the boundaries between the stages for transient staging of data.

{\em Data Access} components are gateways to Data Sources and Data
Storage components, and provide interfaces that often enforce security
controls to restrict access, provide dynamic information about the
available data sources and/or data analysis components, and may also
perform logging of requests and data movement (e.g., tracking
provenance for auditing). A side effect of auditing/logging may be the
growth of provenance metadata simply due to access transactions even
if no new data is created. Data Access components usually serve at the interface between the application stages, before
and after data movement.

{\em Data Movement} components are used for bulk or continuous
movement of data between parts of the application, or to and from the
application, %
and visible to the application (e.g., staging), or implicit (e.g.,
automatic replication of files). Components may operate on streams or
blocks (files). Dynamic characteristics include partitioning,
scheduling and planning (e.g., edge caching of ``hot'' data, where
datasets that are being used frequently during some period are
replicated close to the application). Data movement can be considered
to be real-time (i.e., as it happens, e.g., sensor data), `as real-time'
(e.g., a storage buffer has been introduced into the transport system,
but the data source is still real-time, as with time-series data), or
asynchronous (where the ordering of the data movement is independent
of the data itself).  Latency may be considered as an important
characteristic of the data movement infrastructure depending on the
application, e.g., real-time processing of sensor data for disaster
analysis. Other dynamic aspects of data movement include data rates,
their variability from data sources, and the ``freshness'' of
data. These infrastructure appear on the edges between the application
stages.

{\em Data Discovery} components assist the access to and use of data
sources, and include Catalogs, Information Services, and Metadata
stores. The dynamic characteristics pertain to the evolution or
recalibration of information held in these components. These components are orthogonal to the
applications stages and can be used by any of the
stages to guide their selection or identification of incoming data sources and outgoing data sinks
or storage.

{\em Notification} components respond to changes in the application or
data and trigger other functionality. These include publish-subscribe message brokers,
and complex event processing. By nature, these components support the dynamic aspects of an application.  These components again are orthogonal to the applications stages and can be used for coordination of the stages and data movement.

In each case, these components can be single or centralized,
distributed, partitioned or federated. These are also typically
offered as middleware and platform services, or system-level software
capabilities.

\subsection{Infrastructure to Support Data Analysis}
\label{sec:infra_analysis}

Analysis components include functionality such as comparison,
statistical analysis, event analysis (i.e., analyzing information
derived from data rather than raw data) and visualization. These components leverage
the computational aspects of the infrastructure.
Based on the application stages outlined in
\figurename~\ref{fig:figures_application-stages}, we identify a number
of infrastructure services for data analysis that can be associated with these stages.

{\em Conversion} components (associated with the Data Transformation
and Decisions stage) deal with the changing of data into a different
form, including file conversion, (re-)formatting, transformation and
metadata extraction. Heterogeneity of formats and representations is
common in many scientific disciplines and such components often
provide essential pre-processing of data prior to analysis. These are often
referred to as adaptor services (or ``shim'' services in the Taverna workflow system~\cite{Taverna}).

{\em Enrichment} components take data and attach additional detail to
them, which might include filtering, image processing, information
integration, data tagging, semantic annotation, data augmentation, or
data fusion (e.g., OLAP). Many of these components may deal with
distributed sources of data. It is useful to note that these
operations may not be fully automated and may involve a human
expert. These components would be associated with the Data
Transformation and Decisions stage.

{\em Analytics and visualization} components help extract domain knowledge and assist with
exploration of the data. These are often closely supervised by the end-user and may even be
interactive. The proliferation of ``big data'' has caused data mining libraries, machine-learning
toolboxes and visualization environments to become part of the infrastructure, some of which are
deployed for specific domains but others which are more generic (e.g., Weka, GraphLab, Apache Mahout). These components are associated
with the Data Processing stage.

\subsection{Infrastructure to Support Coordination within Applications}
\label{sec:infra_coord}

Programming abstractions and orchestration frameworks cut across both data
management and analysis functions. While different in character from the other
infrastructure discussed in this section, they enable the coordination of the
different stages of the application, sometimes enforcing a particular constraint to improve, e.g., scalability. These approaches can allow the composition of sequential data processing pipelines and further offer the flexibility of modeling more complex constructs.
Standardization of APIs and components within these frameworks can allow
interchangeability of components with similar functionality, such as those described in \S\ref{sec:infra_mgt} and \S\ref{sec:infra_analysis}.

{\em Orchestration} components handle the management and
choreography of data placement, code placement, communication, and
messaging to allow complex analyses to be performed on distributed
data sources. These are often composed as workflows or Directed Acyclic Graphs (DAGs) that capture
the execution dependency between
components. They may also be dynamic if they have the capability to
adapt their behavior based on earlier parts of the workflow or can be
steered based on external notification received during execution. In
addition to supporting choreography, while often making use of one or more
resource manager/scheduler components, an orchestration component may also
include specialist support for conversion and analysis components
discussed above.

{\em Bulk Data-parallel} frameworks use ``Bulk'' operators to express
computations over collections of data which are executed in a loosely-coupled,
data-parallel manner.  The {\em MapReduce} paradigm~\cite{MapReduce} and its
variants~\cite{twister,hmr} are the most prominent among these, with those like
Pregel and Apache Giraph~\cite{pregel} supporting specific domains like graph
analytics. MapReduce forces programmers to model computation in two stages: a
\emph{map} phase in which a transformation is applied to every element of a
collection independently of every other; and a \emph{reduce} phase in which the
transformed results are aggregated through a binary function that is associative
and commutative.  These properties serve to guarantee that the MapReduce
computation can be efficiently parallelized, by performing the map phase
concurrently and then reducing without the reduction order affecting the final
result.

{\em Stream processing} (sometimes called event processing) is an approach modeled
on signal processing, in which a stream of data is transformed ``on-the-fly'' to produce one or
more further streams of values. As such it is a programming
abstraction suitable for continuous data feeds, or when results need
to be pushed to users of interactive applications. Since the data is
not assumed to be available \textit{en masse}, it can be used in
situations where storing the entire dataset is undesirable,
unnecessary, or impossible.

{\em Sensor networks} and other intelligent front-end data collection systems are sometimes seen as an extension of stream processing and application pipelines.
They often have the feature of being \emph{adaptive}, in that they can vary their exact behavior according either to external control stimuli or from observations made of the data they are
themselves observing. Adaptation may also be through users interactively changing the
the pipeline, either by changing parameters or by adding or removing
tasks in the pipeline. This allows decision logic to be initiated by
the pipeline itself, or using an external control signal that may come
from users or an instrument. This form of ``computational steering''
is especially useful in systems being used to explore a dataset. In these cases, it is important to
note that the infrastructure is adaptive, rather than the scenario itself.

\subsection{Examples of Specific Cyberinfrastructure Usage}

We conclude this section by revisiting the application scenarios of
\S\ref{sec:scenarios} in terms of the infrastructure and coordination approaches that they use
currently, and discuss elements that could usefully be deployed.

The software infrastructure components we have classified above
address different characteristics of D3
applications. \tablename~\ref{tab:inf_app_mapping} summarizes the role of
these components in addressing dynamic and distributed characteristics
of the D3 applications in \S\ref{sec:scenarios}. Each application makes use of different components, though it can be seen that applications that utilize stream or event processing %
have larger requirements on notification and orchestration components.

\begin{table}[h]
  \begin{scriptsize}
    \begin{center}
      \caption{Infrastructure used to support dynamic and distributed
        properties of application scenarios %
      }
      \label{tab:inf_app_mapping}
      \begin{tabular}{|p{6.7cm}|p{6.7cm}|}

\hline	
\multicolumn{2}{|l|}{{\textbf{Application Name}}} \\ \hline	
Infrastructure to support Dynamic Properties & Infrastructure to support Distributed Properties
 \\ \hline
		\hline	
\multicolumn{2}{|l|}{{\textbf{Next Generation Sequencing (NGS) Analytics} (\S\ref{bioSilvia})}} \\ \hline	
Dynamic workload decomposition and distribution for resource optimization \emph{(Orchestration)}
		&
Distributing data from source (generation) to the computing destination (analysis) \emph{(Data
  Discovery, Movement)}
\\
		\hline
\multicolumn{2}{|l|}{{\textbf{ATLAS (an example of use of the WLCG) Reconstruction app} (\S\ref{WLCGSteve})}} \\ \hline
Manage gradual growth of data using SRM \emph{(Data Storage, Access)}
&
Management, subset replication, data movement and partitioning using gLite and SRM \emph{(Data
  Storage, Access, Movement, Discovery)}
              \\
		\hline
\multicolumn{2}{|l|}{{\textbf{LSST} (\S\ref{astroAdam})}} \\ \hline	
	
Steering observations, Databases to manage data growth \emph{(Notification, Data Storage)}
            &
Coordinate classification, distributed databases, data movement to distributed observatories
\emph{(Notification, Data Access, Data Movement, Data Storage)} %
\\
		\hline
\multicolumn{2}{|l|}{{\textbf{SOA Astronomy Applications} (\S\ref{astroSOAAdam})}} \\ \hline	
	
Catalog contents updated, Query mix changes \emph{(Data Discovery)}
             &
Distributed VO Services for data storage and access \emph{(Data Discovery, Data Access, Data Movement)}
\\
		\hline
\multicolumn{2}{|l|}{{\textbf{Cosmic Microwave Background} (\S\ref{astroJulian})}} \\ \hline	
	      Dynamic resource selection  \emph{(Data Discovery)}
& Compute on distributed HPC centers \emph{(Data Movement)}
\\
		\hline
\multicolumn{2}{|l|}{{\textbf{Sensor Network Application} (\S\ref{sensorSimon})}} \\ \hline	
	     Notification of sensor data when in range, data rates and volumes \emph{(Data Movement, Notification)} %
&Aggregating distributed sensor data \emph{(Data Movement, Notification)}  \\
		\hline
\multicolumn{2}{|l|}{{\textbf{Climate (ESGF)} (\S\ref{climateDan})}} \\ \hline	
	      Manage data growth %
             &
Distributed, replicated catalogs using ESGF software stack (GridFTP, wget/HTTP) \emph{(Data Access,
  Data Discovery, Data Movement, Data Conversion)} \\
		\hline
\multicolumn{2}{|l|}{{\textbf{Interactive Exploration of Environmental Data} (\S\ref{envJon})}} \\ \hline	
On-demand data subsetting, caching and versioning, real-time instrument data \emph{(Data Discovery, Data
  Movement, Data Sources)}
		&
Distributed Data Services, Access Restrictions \emph{(Data
  Movement, Data Access)}

\\
		\hline
\multicolumn{2}{|l|}{{\textbf{PowerGrids} (\S\ref{power})}} \\ \hline	
	
Stream processing over variable sources, rates and volumes, Event processing over different
patterns, Trigger analysis on-demand  \emph{(Data
  Sources, Notification, Enrichment,  Orchestration)} %
&
Dispersed event sources and sinks, Launch ensemble modeling on distributed clusters and cloud
\emph{(Data Sources, Notification,  Orchestration)} %

 \\

		\hline
\multicolumn{2}{|l|}{{\textbf{Fusion (ITER)} (\S\ref{fusionScott})}} \\ \hline	
	     Stream and transform data between simulation sites on-demand \emph{(Data Movement, Enrichment)}
	    &
Run analysis on available, distributed resources \emph{(Data Movement)} %
            \\
		\hline
\multicolumn{2}{|l|}{{\textbf{Industrial Incident Notification and Response} (\S\ref{Kees})}} \\ \hline	
	     Acquire different data sources and rates, Processing determined by data
             \emph{(Data Sources, Data Access, Notification, Orchestration)}   %
&
Distributed data sources, Analysis by agents across institutions, Access control \emph{(Data Sources, Data
  Access)} %
            \\
		\hline
\multicolumn{2}{|l|}{{\textbf{MODIS Data Processing} (\S\ref{modisKeith})}} \\ \hline	

--

		&
Data from distributed FTP sources, Cloud queues for coordination, Partitioned NoSQL data \emph{(Data
  movement, Data Storage, Orchestration)}
\\
		\hline

\multicolumn{2}{|l|}{{\textbf{Distributed Network Intrusion Detection} (\S\ref{detectionJon})}} \\ \hline	
Analyze dynamic network traffic rates \emph{(Data Movement)} %
            &
Run analysis at different sites, Distribute generated data \emph{(Data Movement)} %
\\
		\hline

      \end{tabular}
    \end{center}
  \end{scriptsize}
\end{table}

\tablename~\ref{table:progSystems} summarizes the coordination approaches
currently dominant in each application case study. It is worth noting
that, in common with most long-lived systems, many of the cases
have accreted substantial code bases over time and so show little in
the way of systematic architecture. (One reason for the dominance of
workflow pipelines may be that such architectures are quite suitable for adding
extra processing steps to existing computations.)

\begin{table}
  \begin{scriptsize}
    \begin{center}
      \caption{Coordination approaches used by the applications\label{table:progSystems}}
      \begin{tabular}{|p{6.3cm}|p{7.2cm}|}
        \hline Application  & Current approaches \\
        \hline
        \hline NGS Analytics & pipelines that ship with
        sequencers~\cite{illumina}; workflows~\cite{Taverna} \\
        \hline ATLAS & custom algorithms; job submission tied to WLCG \\
        \hline LSST & custom layer; scripted workflows; queries \\
        \hline SOA Astronomy & web services~\cite{barker} \\
        \hline Cosmic Microwave Background & \textit{ad hoc} \\
        \hline Sensor Network Application & focused on data collection \\
        \hline Climate & \textit{ad hoc} \\
        \hline Interactive Exploration of Environmental Data &
        visualization tools~\cite{blower2} \\
        \hline Power Grids & stream processing~\cite{Simmhan:sciencecloud:2011}; data mining~\cite{Yin:mapreduce:2012} \\
        \hline Fusion (ITER) & scripted pipeline~\cite{adios} \\
        \hline Industrial Incident Notification and Response &
        agent-based~\cite{emergency4}  \\
        \hline MODIS Data Processing & scripted pipeline~\cite{modis:ipdps:2010} \\
        \hline Distributed Network Intrusion Detection & monolithic
        packages \\
        \hline
      \end{tabular}
    \end{center}
  \end{scriptsize}
\end{table}

The applications in \tablename~\ref{table:progSystems}, which make use
of scripted or \textit{ad hoc} pipelines, could in most cases be easily
replaced with a structured workflow engine. Doing so would bring
immediate benefits in terms of reproducibility, but also in terms of
standardization of components and interfaces.

A further implication of standardization is that
component reuse is greatly improved. This is perhaps best illustrated
by observing the increased ability to automate the processing step
over repeated runs.  The use of standard component-based software
techniques such as web services, coupled with web technologies for
metadata representation, assist in reuse, re-purposing and
integration, for instance into different workflow systems.

The {\em MapReduce} paradigm~\cite{MapReduce} and its variants~\cite{twister,hmr} are the most prominent among the bulk data-parallel frameworks which use bulk operators to express computations over collections of data in an loosely-coupled manner.  Applications such as Next Generation Sequencing (NGS) (\S\ref{bioSilvia}) and Power Grids (\S\ref{power}) use MapReduce for
their analytics~\cite{cloudburst,genometoolkit,Yin:mapreduce:2012}.

There are many examples of stream~\cite{chandrasekaran:telegraphcq2003,
abadi:borealis2005,
liew:streamgraph2010} and event~\cite{ESPER, siddhi,
MicrosoftInsight} processing frameworks, and recent works have
combined streaming with more traditional file-based
workflows~\cite{ZinnCluster,Herath:CCGrid:2010}.  Applications such as
the Marine Sensor Network (\S\ref{sensorSimon}), Distributed Network
Intrusion Detection (\S\ref{detectionJon}), Power Grids
(\S\ref{power}), and Industrial Incident Notification and Response
(\S\ref{Kees}) use stream and event processing abstractions.

The view of scientific applications as pipelines hides significant
differences between (and within) domains, since the different stages
of the pipeline receive different emphases in different
applications. Some applications must collect and store all the
incoming data, with only minor pre-processing to clean the data ahead
of the main processing. For others, a major constraint is to reduce
the incoming data stream as quickly as possible, perhaps discarding a
large fraction of the data before any processing occurs. In spite of these differences, we believe the use of the pipeline abstraction is still useful.
Taking some of the examples from
\S\ref{sec:scenarios}, medical imaging and astrophysics often fall
into the first category, while particle physics (especially in the
case of the Large Hadron Collider) often falls dramatically into the
second, to the extent that principled data reduction forms one of the
major scientific challenges for the project.

There are also significant variations in where ``the pipeline'' begins
and ends. This represents the difficulty in differentiating between ``traditional'' applications
and ``infrastructure'' applications. Many systems are driven by datasets collected and shared
widely, against which several different applications may be applied.
Here the infrastructure application is the collection of components used to allow the
datasets to be accessed and analyzed by the different applications.
Climate science is a good example of this approach. In this
case the data is ``pre-collected'' in the sense of existing outside
most of the applications, although its collection may have been a
complex task. The raw data required for gene sequencing
(\S\ref{bioSilvia}), for example, while being collected using a
dedicated and very sophisticated computerized processing system, would
generally be regarded as ``given'' and not part of the processing
pipeline, since the pipeline would not affect the collection
task. Another way to look at this is that data is collected
\emph{independently} of its further processing.

Alternatively the application may be driven directly
from a live data stream, with processing happening on the
fly. The front-end data collection subsystem is becoming increasingly
flexible and capable, with the introduction of sensor networks to
replace more passive approaches, and it may often make sense to
include the sensor network into the application workflow pipeline, such as for Power Grids and
Industrial Incident Response. All
these choices have implications for design and provisioning. For instance,
the choice of which sort of pipeline a particular application uses may
well be fixed, or will change only slowly as the experimental
environment evolves. For example, we see some systems that add simulation alongside experimental data collection to act
as a verification and validation step.

Revisiting the overall issue of infrastructure, which we defined as
the hardware and software needed to run the applications, we
note that the specific hardware is not a primary issue.  The
applications we have discussed all use a mix of the systems that exist
today, which have aspects of web, grid, and cloud, and can be
considered a hybrid of the three.  The key infrastructure components
are really the software and services on top of the hardware.  In this
section, we have discussed infrastructure to support data management and
data analysis, as well as the tools and systems used to program
applications, and the specific components used for the application in
\S\ref{sec:scenarios}.  However, there is not much commonality in the
software and services that the applications use, so the software and programming infrastructure
we have described here is much more broad than what is used by the
individual applications we have studied.

\section{Discussion and Conclusions} \label{sec:conclusions}

This section brings together some key observations from the analysis of
applications and infrastructure carried out in this paper.
Section~\ref{sec:discussion} provides a discussion of dynamic and distributed
characteristics of data associated with a set of applications.

We suggest two metrics that may be associated with such applications, primarily
as a basis to compare how compute and data processing (in terms of their number
and distribution) may be used to find common features between them. Such feature
analysis is useful, we believe, for two main reasons: (i) to identify
similarities between the applications covered in \S\ref{sec:scenarios} with
others that are in the design phase or under development, so that common
requirements can be identified; and (ii) to identify gaps in current software
tools and programming systems that such applications use, thereby leading to the
development of new tools and systems.

As our application and infrastructure coverage is, by necessity, limited,
identifying common features also enables the lessons learned to be generalized
and applied more widely. We subsequently discuss a set of architectures for D3
science, and pose some questions we believe are pertinent as the community
begins considering the next-generation architectures for D3 applications. In
\S\ref{sec:conclusion} we conclude by providing a number of observations to
inform applications and cyberinfrastructure developers, based on common trends
found in our study.

\subsection{Discussion}
\label{sec:discussion}

We summarize four key themes from previous sections: (i) the source and nature
of dynamic data observed in the applications surveyed in \S\ref{sec:scenarios};
(ii) the types of data distribution observed in such applications; (iii) the
general characteristics that can be associated with D3 science applications;
(iv) the characteristics of a computational infrastructure most suited for D3
science applications.  Whereas (i) and (ii) are observations that can be made on
existing applications discussed in \S\ref{sec:scenarios}, (iii) and (iv) attempt
to synthesize general trends that can be applied to applications we have not
considered.

\subsubsection{Understanding Dynamic Data }

The applications in \S\ref{sec:scenarios} show that the dynamic characteristics
of many applications arise in two ways: either due to data generation and
placement, or due to the certain aspects of the application.  The first
of these refers to data dynamism and the second to application dynamism
(outlined in \S\ref{sec:dynTypes}).

\emph{Data generation, placement and consumption}: In many
applications, data is dynamically generated in response to a user
request or when a particular phenomenon of interest has been
observed. Often this data is large in size and must be placed close to
the analysis. In some instances, it may not be
known where (and when) an analysis request may be made, and it is
therefore necessary to identify how replicas of the data could be
produced and placed across multiple possible locations.  In general,
dynamic data arises due to changing data properties, changing data
volume (generation, consumption, or processing), and changing data
location or distribution.

\emph{Application dynamism: } In many scientific applications the
workflow structure is well defined, but either the components used
within this structure or the location of the data that feeds the
workflow may change dynamically.  Data coordination and planning
(identifying how data is managed from ingest through analysis to
output) within such applications can range from statically defined
interactions between execution units and data sources to a fully
adaptable interaction.  As outlined in the
Industrial Incident Notification and Response application in
(\S\ref{Kees}), the workflow structure may also change based on a user
request or the event being observed. Data relating to the coordination
of a components that make up the application can therefore also be
dynamic and change.  Exposing the data planning used within an
application makes it possible to modify or adapt it over time.

\subsubsection{Understanding Distributed Data}

\begin{figure}[ht]
	\centering
		\includegraphics[width=.7\textwidth]{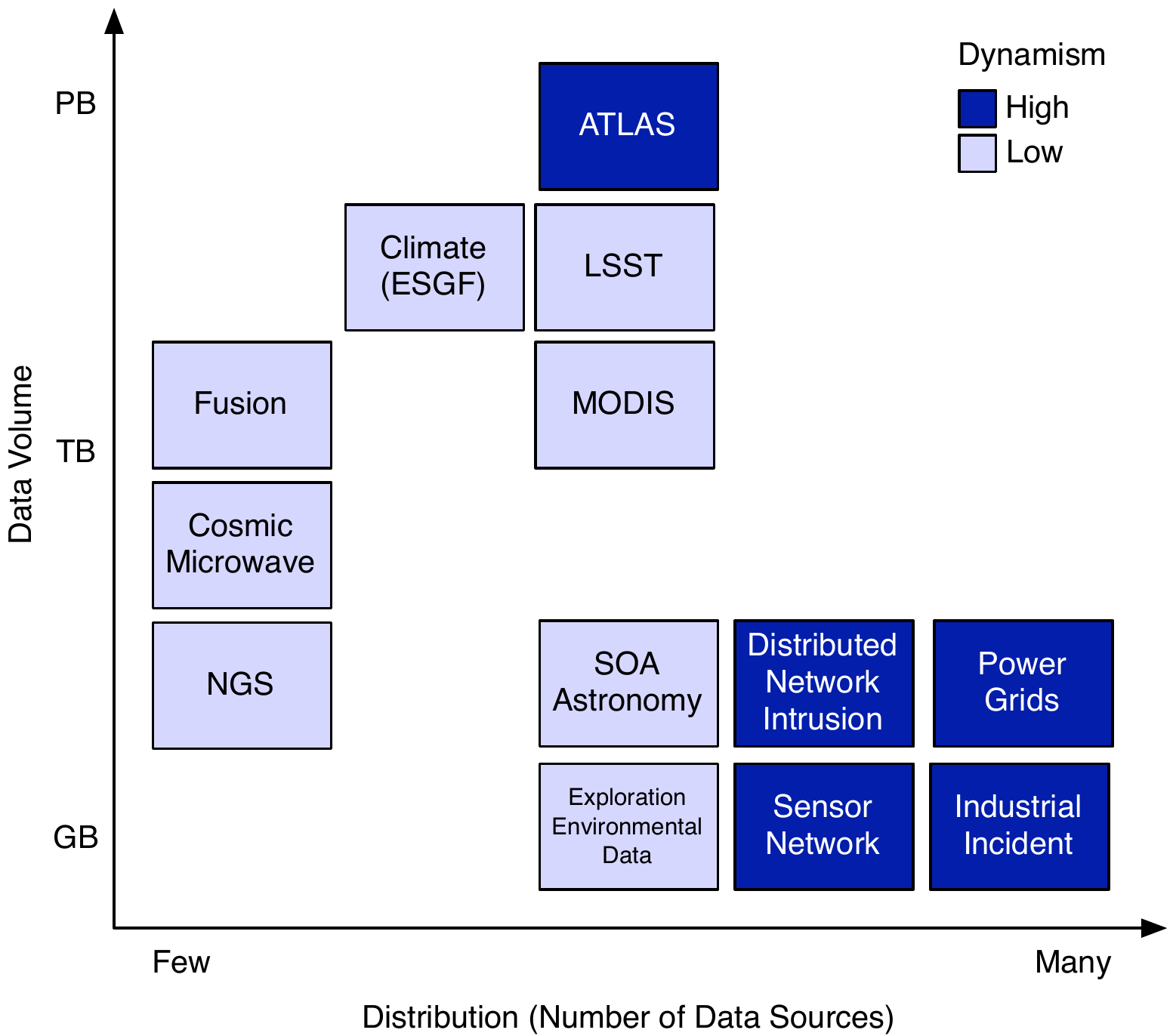}
                \caption{The distribution, dynamism, data volume
                  characteristics of the surveyed D3 applications. The
                  clustering of applications with high dynamism in the
                  bottom right hand corner indicates a similar
                  structure amongst these applications, namely
                  multiple (sensor-based) sources of data
                  generation with temporal variation in data
                  generation. Although we do not explicitly
                  color/shade traditional and infrastructural
                  applications, it is worth mentioning that there is
                  no correlation between application type and any of
                  the three ``D''s; we attribute this to a strong
                  dependence of degree-of-distribution infrastructure
                  available for both types of applications.
                  \label{fig:figures_application}}
\end{figure}

From \S\ref{sec:distTypes} and \figurename~\ref{fig:figures_application}, we
find that applications with distributed data vary in the amount or volume of
data distributed, the degree of distribution (scattered over one site or many
sites), and the dynamic aspects associated with distribution, e.g., latency and
relative scheduling flexibility (and thus tolerance or constraints on data
transfer/aggregation).

Some applications use specialized data cyberinfrastructure, while some run
utilize existing HPC environments.  On the other end of the spectrum are highly
dynamic applications that process sensor data and make near-realtime decisions
based on streaming data, e.g., the power grid and sensor network applications.

{\it Data topology} can be used to identify the flow of data through the
application, with nodes representing execution units and edges representing
communication features.  One aspect of managing data topology is ``data
planning," which refers to how the data is coordinated from ingest through
analysis to output. In all cases, there is a coordination of the data with code
used to analyze or transform it. In some cases these form clear, integrated
stages where data flows from one to another. In others the way that data enters
and flows through the application is more loosely defined and the execution of
code is independent. Some examples include the ATLAS experiments, which have a
single point of primary data generation which fans out to multiple analysis
sites followed by mass exchange of derived data, and sensor grids, which have
multiple sources of data generation collected to a single point of analysis
followed by constrained point of reuse.

Some applications must collect and store all incoming data, with only minor
pre-processing to clean the data ahead of the main processing. For others, a
major constraint is to reduce the incoming data stream as quickly as possible,
perhaps discarding a large fraction of the data before any processing occurs.

Although dynamic and distributed properties have been characterized
separately, it is important to remember, as the above patterns
indicate, that distribution and dynamism become increasingly
correlated at extreme scales.  For example, applications that run on
distributed infrastructures are often also highly dynamic and vice
versa.

\subsubsection{Understanding Distributed Dynamic Applications }
\label{sec:quantifyingdistapps}

In \S\ref{sec:distdyndata} we characterized dynamic and distributed
properties of the D3 Science applications. Our analysis was
qualitative by necessity, due to the fact that approaches to quantify
dynamism and distributed properties were often intimately tied to
application/scenario definition and specific usage modes.
Notwithstanding, to quantify better dynamic properties of an
application, we believe a term analogous to the use of Amdahl's
number~\cite{10.1109/ICDE.2000.839382}, such as [Number\ of
Compute\ Operation / Amount\ Data (Consumed,\ Generated\ or\ Transformed)]
could be used to characterize `static' data-intensive applications and systems.
Note that this term is scale-invariant, i.e., a value of 1 could
correspond to very small or very large values of both numerator and
denominator, and that it does not capture the time scales over
which data changes.

Similarly, another term [Amount of Data Transferred / Degree of Distribution]
could be used to quantify the
``scale'' of distribution, namely the degree of distribution as a
function of the amount of data transferred, where the degree of
distribution can be defined as the number of distinct data locations.

It has proven difficult to provide ``global'' quantitative values of
these ratios, as often the value of the numerators is
subject to how an application/scenario is defined, even more so for a
``class of applications'' as opposed to a specific application. We
believe, nonetheless, that these measures 
suggest a possible way to quantify an application's dynamism and distribution 
and thereby find possible common approaches solutions.  We plan to explore this 
in our future work, and also hope others will consider building on these
initial concepts.

\subsubsection{Towards an Architecture for D3 Science?}

An investigation of data-intensive distributed applications on production
cyberinfrastructure, leads us to believe that there are three primary
macroscopic architectures and approaches for distributed data-intensive
applications: (i) localize all data into a ``cloud'' (or a large analytical
engine) -- the paradigm adopted by several genome projects and the cancer atlas,
(ii) decompose and distribute data to an appropriate number of
computing/analytical engines as available -- the paradigm employed by particle
physics for the discovery of the Higgs Boson (similar to the application in
\S\ref{WLCGSteve}), and (iii) a hybrid of the above two paradigms, wherein data
is decomposed and committed to several infrastructure, which in turn could be a
combination of either of the first two paradigms, similar to what is done in the
Power grids application
(\S\ref{power}). %

It is obvious that the first architecture can be used for large volumes of data,
but there are self-evident limitations to the scalability or validity of this
model. How does this limitation change, qualitatively or quantitatively, if
dynamic data and/or applications are considered? This begets the question: how
fundamental an issue is {\it dynamic}?

The high-energy physics (HEP) community has self-organized towards the
second architectural approach.  Although distributed computing
infrastructures representing this approach, such as OSG and EGI, have
been successful for this community and application, broad uptake
across other application types, modes, and data volumes has not taken
place. What lessons does the HEP experience teach us for the other
communities, such as bioinformatics?  Can an infrastructure
conforming to a particular architecture and developed for a specific
application domain be generalized to another science domain, using the
same underlying macro architecture but by providing a different set of
services and specialized data cyberinfrastructure? Or is the
architecture specific to a {\it fixed} range of compute-data
characteristics? If so, what are these characteristics?

\subsection{Conclusions}
\label{sec:conclusion}

The growth in data volumes and its importance is having profound implications on
the way applications are designed and the way we develop infrastructure and
provision associated services.  Many fields, such as biology, that until
recently were not characterized by their intensive use of data are rapidly
becoming data-driven.

Associated with the growth in data volumes are increasing levels of dynamism and
distribution; sometimes distribution is logical, but sometimes the data is
physically distributed for reasons of performance or convenience. At large
scales, dynamism becomes intrinsic to the application and infrastructure;
sometimes it occurs because of changing experimental or analytic conditions,
sometimes due to the availability of infrastructure and resources.

An important motivation for this work is to capture the state of the art in the
design and development of distributed dynamic applications and the associated
infrastructure currently used for their execution. We have aimed to try to
discern common approaches and challenges, and to identify any obvious gaps (or
opportunities). Using a common vocabulary and terminology to describe otherwise
distinct and unrelated applications allows potential application
users\slash developers to benefit from the insights that a common analytical framework
provides. For example, seeing how applications with similar patterns (e.g.,
pipelines) have addressed issues of dynamism and distribution enables common
solutions and facilitates the emergence of best practices even in the absence of
formal methods.  Similarly, our analysis sends a useful message to domain and
application scientists that the challenges and barriers they face are not unique
even if current solutions require a level of customization.  In this paper, we
have identified many different applications characteristics, programming
systems, and infrastructure techniques that either support dynamic data or
produce it.

In \S\ref{sec:scenarios}, we divided applications into groups: traditional and
infrastructural applications.  Traditional applications have been built by one
or more authors in order to solve a particular problem or to answer a science
question.  Infrastructural applications have been built by groups or communities
to allow the solution of a set of problems, or to answer a set of science
questions. All applications in general, but infrastructural applications in
particular cannot be developed in vacuum; they must be executed on shared
infrastructure, and are thus sensitive to the external tools and services, as
well as their provisioning as software-systems.

As part of this survey, we observe that there exist patterns in the described
applications; however, implementation and deployment specific details can often
mask these patterns.  For example, applications often differ in the way they
handle distribution and dynamism, such as managing such as data/compute
localities. This makes it difficult to support patterns of distribution and
dynamism in a general-purpose fashion, for anything but the simplest patterns.

\section*{Acknowledgements}
The work was sponsored by NSF OCI-1059635 and made possible by the UK
e-Science Institute Research Theme on Dynamic Distributed Data-Intensive
Programming Abstractions and Systems.  SJ contribution was also funded by NSF
CAREER ACI-1253644.  We thank Simon Dobson, Jon Weissman and Jon Blower for
useful initial discussions, as well as 3DPAS workshop attendees. Discussions of
the selected applications benefited from the generous help of Adam Barker, Jon
Blower, Julian Borrill, Silvia Delgado Olabarriaga, Steve Fisher, Keith Jackson,
Scott Klasky, Bob Mann, Don Middleton, Kees Nieuwenhuis, Manish Parashar,
Stephen Pascoe, and Jon Weissman, though any errors are the responsibility of
the authors.  Some of the work by Katz was supported by the National Science
Foundation while working at the Foundation; any opinion, finding, and
conclusions or recommendations expressed in this material are those of the
author(s) and do not necessarily reflect the views of the National Science
Foundation.

\bibliographystyle{unsrt}

\newpage
\appendix
\section*{Appendix: Methodology}

This paper originated in work at the UK e-Science Institute (eSI) at
the University of Edinburgh, in a research theme examining Dynamic
Distributed Data-intensive Programming Abstractions and Systems,
called 3DPAS~\cite{3dpas-theme}.  The description of
this theme's work was:

\begin{quote}
Many problems at the forefront of science, engineering, medicine, and the social sciences, are increasingly complex and interdisciplinary due to the plethora of data sources and computational methods available today. A common feature across many of these problem domains is the amount and diversity of data and computation that must be integrated to yield insights.

For many complex data-intensive applications, moving the data may have restrictions. Increasingly important type of data-intensive applications are data-driven applications. For example, data is increasingly large-scale, distributed arising from sensors, scientific instruments \& simulations. Such data-driven applications will involve computational activities triggered as a consequence of independent data creation; thus it is imperative for an application to be able to respond to unplanned changes in data load or content. Understanding how to support dynamic computations is a fundamental, but currently a critical missing element in data-intensive computing.

The 3DPAS theme seeks to understand the landscape of dynamic, distributed, data-intensive computing: the programming models and abstractions, the run-time and middleware services, and the computational infrastructure. It will analyze existing tools and services, identify missing pieces and new abstractions, and propose practical solutions and best practices.
\end{quote}

The theme held three workshops, and in each, application scientists and
computing technologists were invited to give presentations on their work, and
then to be involved in discussions to organize and understand the presented
materials.  Between the workshops, the theme organizers also drafted outlines of
a report that was a predecessor of this paper, for discussion in the next
workshop.  Overall, there were about 50 workshop attendees.

In order to select applications and gather information about
them, during the various workshop discussions, when an
application that ``felt'' new was mentioned, we asked the person who
mentioned it to provide a written set of answers to the following questions about the
application:

\begin{enumerate}
\item What is the purpose of the application?
\item How is the application used to do this?
\item What infrastructure is used? (including compute, data, network, instruments, etc.)
\item What dynamic data is used in the application?
\begin{enumerate}
\item What are the types of data,
\item What is the size of the dataset(s)?
\end{enumerate}
\item How does the application get the data?
\item What are the time (or quality) constraints on the application?
\item How much diverse data integration is involved?
\item How diverse is the data?
\item Please feel free to also talk about the current state of the
application, if it exists today, and any specific gaps that you know
need to be overcome.
\end{enumerate}

We then used the text of these answers to describe each of the applications (in \S\ref{sec:scenarios}) in terms of:

\begin{itemize}
\item What does the application do? What problem does it solve?
\item How does it do it? How does the application run?  What infrastructure does it need/use?
\item What are the big data aspects of the application?
\item What are the distributed aspects of the application?
\item What are the dynamic aspects  of the application?
\item What else is important about the application?  What does it do well?  Poorly? How was the app written? What tools does it use?
\end{itemize}

After the workshops, the theme organizers and a set of attendees who
were interested in participating started discussing how to organize
and analyze the applications and the technologies, including
infrastructure, programming systems and abstractions.  These
participants are the current authors of this paper, and this paper is
the result of that process.

As part of the theme, two 3DPAS workshops~\cite{3dapas,D3-escience}
were organized at major conferences: HPDC 2011 and IEEE eScience 2011.
These workshops had both peer-reviewed and invited papers.

A combination of invitation-only focused discussion meetings and
broader community engagement events were used to ensure that the
applications surveyed and ideas captured in this paper are
representative of current community thinking as well as of
intellectual relevance.

\end{document}